\documentclass[aps,twocolumn,prl,showpacs,floatfix,superscriptaddress]{revtex4-1}

\usepackage{amsfonts}
\usepackage{amssymb}
\usepackage{amsmath}
\usepackage{latexsym}
\usepackage{array}
\usepackage{dcolumn}
\usepackage{longtable}
\usepackage{colortbl}
\usepackage{bm}
\usepackage{natbib}
\usepackage{graphicx}

\usepackage[T1]{fontenc}
\usepackage[utf8]{inputenc}
\usepackage[ugly]{units}		
\usepackage{siunitx}	
\usepackage{braket}	
\usepackage{booktabs}
\usepackage{listings}
\usepackage{slashed} 

\usepackage[colorlinks=true,linkcolor=black,citecolor=blue,
urlcolor=blue,bookmarksopen]{hyperref}

\begin{document}

\title{Search for the effect of massive bodies on atomic spectra and constraints on Yukawa-type interactions of scalar particles}
\date{\today}

\author{N.~Leefer}
\email[E-mail: ]{naleefer@gmail.com}
\affiliation{Helmholtz-Institut Mainz, Johannes Gutenberg-Universität Mainz, 55128 Mainz, Germany}
\author{A.~Gerhardus}
\email[E-mail: ]{andreas.gerhardus@uni-bonn.de}
\affiliation{Bethe Center for Theoretical Physics, Physikalisches Institut der Universität Bonn, 53115 Bonn, Germany}
\author{D.~Budker}
\affiliation{Helmholtz-Institut Mainz, Johannes Gutenberg-Universität Mainz, 55128 Mainz, Germany}
\affiliation{Physics Department, University of California, Berkeley 94720-7300, USA}
\affiliation{Nuclear Science Division, Lawrence Berkeley National Laboratory, Berkeley, California 94720, USA}
\author{V.~V.~Flambaum}
\affiliation{Helmholtz-Institut Mainz, Johannes Gutenberg-Universität Mainz, 55128 Mainz, Germany}
\affiliation{School of Physics, University of New South Wales, Sydney 2052, Australia}
\author{Y.~V.~Stadnik}
\email[E-mail: ]{y.stadnik@unsw.edu.au}
\affiliation{School of Physics, University of New South Wales, Sydney 2052, Australia}

\begin{abstract}
We propose a new method to search for hypothetical scalar particles that have feeble interactions with Standard-Model particles. 
In the presence of massive bodies, these interactions produce a non-zero Yukawa-type scalar-field magnitude. 
Using radio-frequency spectroscopy data of atomic dysprosium, as well as atomic clock spectroscopy data, we constrain the Yukawa-type interactions of a scalar field with the photon, electron, and nucleons for a range of scalar-particle masses corresponding to length scales $ > 10$~cm. 
In the limit as the scalar-particle mass $m_\phi \to 0$, our derived limits on the Yukawa-type interaction parameters are:~$\Lambda_\gamma \gtrsim 8 \times 10^{19}$~GeV, $\Lambda_e \gtrsim 1.3 \times 10^{19}$~GeV, and $\Lambda_N \gtrsim 6 \times 10^{20}$~GeV. 
Our measurements also constrain combinations of interaction parameters, which cannot otherwise be probed with traditional anomalous-force measurements. 
We suggest further measurements to improve on the current level of sensitivity. 
\end{abstract}


\maketitle

Cosmological and astrophysical observations provide strong evidence for a dark matter- and dark energy-dominated universe~\cite{Hinshaw2013,Planck1502Ade}. 
While the nature of dark matter and dark energy is unknown, the evidence from cosmology and astrophysics has motivated numerous laboratory searches for non-gravitational physics associated with the dark sector~\cite{Essig2013}. 
In the present work, we focus on scalar (spin-0) models that can produce local variation of the fundamental constants in the presence of massive bodies \cite{ELLIS1989}. 

A scalar field $\phi$ may interact with the Standard-Model (SM) sector via the Yukawa-type Lagrangian:
\begin{equation}\label{eq1}
\mathcal{L}_\textrm{int} = -\sum_f \frac{\phi}{\Lambda_f} m_f \bar{f}f + \frac{\phi}{\Lambda_\gamma}\frac{F_{\mu\nu}F^{\mu\nu}}{4} \, ,
\end{equation}
where the first term represents the coupling of the scalar field to the SM fermion fields $f$, with $m_f$ the standard mass of the fermion and $\bar{f}=f^\dagger \gamma^0$, and the second term represents the coupling of the scalar field to the electromagnetic field tensor $F$. 
Here $\Lambda_f$ and $\Lambda_\gamma$ are effective new-physics energy scales that determine the relevant non-gravitational coupling strengths. 
Unless explicitly stated otherwise, we adopt the natural units $\hbar = c = 1$ in the present work. 

Comparing the interaction terms in Eq.~(\ref{eq1}) with the relevant terms in the SM Lagrangian, $\mathcal{L}_{\textrm{SM}} \supset -\sum_f  m_f \bar{f}f - F_{\mu\nu}F^{\mu\nu}/4$, we see that the fermion masses and the electromagnetic fine-structure constant $\alpha$ are altered according to (see, e.g., \cite{Stadnik2015DM-VFCs} for more details):
\begin{equation}
\label{eq3}
m_f \rightarrow m_f \left( 1 + \frac{\phi}{\Lambda_f}\right) , ~\alpha \rightarrow \frac{\alpha}{1 - \phi/\Lambda_\gamma} \simeq \alpha \left( 1 + \frac{\phi}{\Lambda_\gamma}\right)  \, . 
\end{equation}

Solving the Euler-Lagrange equation for the full Lagrangian of $\phi$, which includes the kinetic term, $ (\partial_\mu \phi) (\partial^\mu\phi) / 2$, and potential term, $- V(\phi) = - m_\phi^2 \phi^2 / 2$, where $m_\phi$ is the mass of the scalar particle, gives the following equation of motion for $\phi$:
\begin{equation}
\label{phi_EOM}
\left( \partial_\mu \partial^\mu + m_\phi^2 \right) \phi = -\sum_f \frac{m_f \bar{f}f}{\Lambda_f} + \frac{F_{\mu\nu}F^{\mu\nu}}{4\Lambda_\gamma} \, ,
\end{equation}
which shows that SM fermion and electromagnetic fields, in the presence of the interactions (\ref{eq1}), act as sources of the scalar field $\phi$. 
The source bodies that we consider in the present work are composed of atoms, which are composite systems consisting of neutrons, protons, electrons, and strong and electromagnetic binding energies. 
It is, therefore, convenient to express the right-hand side of Eq.~(\ref{phi_EOM}) in terms of the fermion mass-energy and nuclear Coulomb energy densities as $-\sum_{f=n,p,e} \rho_f / \Lambda_f - \rho_{\textrm{Coulomb}} / \Lambda_\gamma$, and so the resulting scalar field generated by a neutral source atom is given by
\begin{widetext}
\begin{equation}
\label{eq4}
\phi(r) \approx - \underbrace{ m_N \left\{\frac{A-Z}{\Lambda_n} + Z\left[\frac{1}{\Lambda_p}+\frac{5\times10^{-4}}{\Lambda_e}\right] + \frac{1}{m_N \Lambda_\gamma} \left[ \frac{a_C Z (Z-1)}{A^{1/3}} + Za_p + (A-Z)a_n  \right] \right\} }_\beta \frac{ e^{-m_\phi r}}{4\pi r} \, ,
\end{equation}
\end{widetext}
where $A$ is the total nucleon number of the nucleus, $Z$ is the proton number of the nucleus, and $m_N = (m_p + m_n)/2 = 0.94$~GeV is the averaged nucleon mass. 
The energy associated with the electrostatic repulsion between protons in a spherical nucleus of uniform electric-charge density, $a_C Z (Z-1) / A^{1/3}$ with $a_C \approx 0.7$~MeV, comes from the Bethe-Weizsäcker formula~\cite{Weizsacker1935}, while the electromagnetic energies of the proton and neutron, $a_p \approx +0.63$~MeV and $a_n \approx -0.13$~MeV, are derived from the application of the Cottingham formula \cite{COTTINGHAM1963} to electron-proton scattering \cite{GASSER1982}.

According to Eq.~\eqref{eq3}, the generated scalar field \eqref{eq4} will result in a modification of the fundamental constants in the vicinity of a massive body. 
Therefore, experiments, which search for possible variations of the fundamental constants (see, e.g., Refs.~\cite{Rosenband2008,Webb2011,Guena2012,Leefer2013a,Berengut2013,Huntemann2014,Godun2014}), can be used as sensitive probes of such scalar fields.

Additionally, the exchange of virtual $\phi$ quanta between two massive bodies results in an anomalous force between the bodies. 
For two point masses $M_1$ and $M_2$ separated by a distance $r$, the anomalous force is described by the potential
\begin{equation}\label{eq5}
V(r) = - \frac{\beta_1 M_1}{\mu_1} \frac{\beta_2 M_2}{\mu_2} \frac{e^{-m_\phi r}}{4\pi r} \, ,
\end{equation}
where $\beta_{1}$ and $\beta_{2}$, defined in Eq.~\eqref{eq4}, depend on the composition of the respective objects, and $\mu_1$ and $\mu_2$ are the nuclear masses for each object. 
Thus, experiments, which search for anomalous forces, including torsion pendulum experiments~\cite{Eotvos1922,Princeton1964,Moscow1972,Adelberger1999_U238,Schlamminger2008,Adelberger2009,Wagner2012torsion}, lunar laser ranging measurements \cite{LLR1988,LLR1996,LLR2004} and atom interferometry experiments \cite{AI2004,Hohensee2013AI,AI2013,AI2014Rasel,AI2014Tino,Tino2014atom,AI2015,Hamilton2015AI}, also serve as sensitive probes of the interactions considered in the present work.

In this letter, we report constraints on the interaction parameters in Eq.~\eqref{eq1} based on radio-frequency spectroscopy of dysprosium and atomic clock measurements. 
The attractive feature of using a spectroscopy-based approach in this context is that spectroscopy measurements require dealing with only scalar quantities, namely the ratio of two transition frequencies at two different distances from a massive body, while traditional anomalous-force measurements usually involve vector quantities, such as the difference in acceleration of two test bodies in the presence of a massive body. 
Additionally, we show that spectroscopy measurements can be used to probe combinations of interaction parameters, which cannot otherwise be probed with traditional anomalous-force measurements. 

The experimental details of atomic dysprosium spectrscopy have been recounted in earlier publications~\cite{Budker1994, Leefer2013a, Hohensee2013, VanTilburg2015}. 
We briefly revisit the main points here. Dysprosium (Dy) is a lanthanide element with $Z = 66$ and seven stable isotopes of mass number $A = 156,158,160,161,162,163,164$. 
This atom has a nearly-degenerate pair of excited, opposite-parity electronic states, conventionally referred to as states $A$ and $B$ (see Fig.~1 of Ref.~\cite{VanTilburg2015}). 
The degeneracy occurs due to a combination of the large relativistic energy level shifts that are common in heavy elements and the complex level structure arising from the incompletely-filled $f$-shell in Dy. 
The consequences, as they pertain to the present work, are: i) a transition that might otherwise appear at optical frequencies 
instead appears at radio frequencies ($< 1$~GHz); ii) the large relativistic corrections make the energy separation sensitive to changes in $\alpha$. 
The combination of i) and ii) results in a sensitive probe of varying $\alpha$ in a system that requires only modest measurement precision. 

The sensitivity of an atomic transition frequency to changes in $\alpha$ can be parametrized as 
\begin{equation}\label{eq6}
\Delta \nu = K_\alpha \frac{\Delta\alpha}{\alpha},
\end{equation}
where $K_\alpha$ is the absolute sensitivity parameter for a given transition with frequency $\nu$. 
For the Dy transition between nearly degenerate levels, $K_\alpha \approx 2\times10^{15}$~Hz~\cite{Dzuba1999a,Dzuba1999,Dzuba2003,Dzuba2008Dy}.

\begin{figure}[t]
  \includegraphics[width=8.5cm]{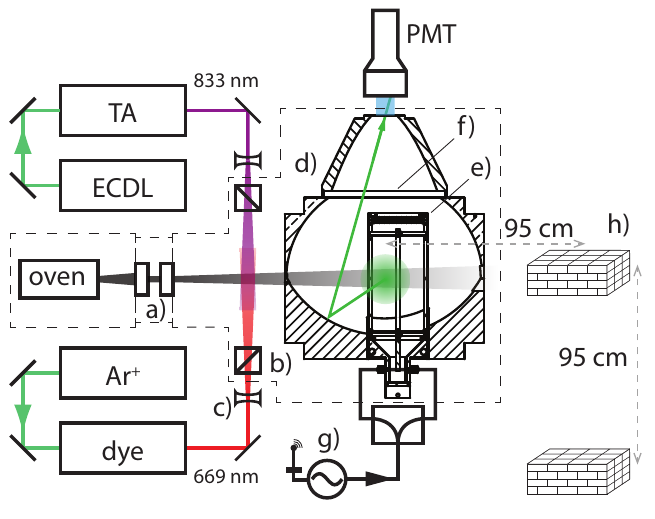}
  \caption{\label{fig1}(Color online) Schematic of the experimental setup (not to scale). 
\textbf{a)} An atomic beam of dysprosium is collimated before entering the interaction chamber. Optical excitation for state preparation is performed with an extended cavity diode laser (ECDL) and a dye laser pumped by an argon-ion laser. 
\textbf{b,c)} Both laser beams are linearly polarized by in-vacuum optical elements. 
\textbf{e)} Within the interaction region, atoms are driven from state $B$ to $A$ by a radio-frequency electric field resonant with the $B\rightarrow A$ transition. 
\textbf{d)} Mirrors direct fluorescence at $\unit[564]{nm}$ to a photomultiplier tube that is placed behind an \textbf{f)} interference filter. 
\textbf{g)} The experiment's frequency reference is provided by a commercial Cs clock. 
\textbf{h)} A 300 kg lead mass can be positioned at varying distances from the atoms by vertical translation. See the text for further details.}
\end{figure}

\begin{table*}[t]
\centering
\caption{Summary of source body parameters, and atomic dysprosium transition frequency variation constraints. 
The results here can be combined with Eq.~\eqref{eq8} to give constraints on the new-physics energy scales that appear in Eq.~\eqref{eq1}, as a function of the scalar-particle mass $m_\phi$. 
We have assumed that the elemental composition of the Sun is $75 \%$ $^{1}$H and $25 \%$ $^{4}$He by mass, and that the elemental composition of the Moon is a 1:1 ratio of $^{24}$Mg$^{16}$O and $^{28}$Si$^{16}$O$_2$ by number. 
}
\label{table1}
\begin{tabular}{ c|c|c|c|c|c| c | c }%
Source & $\beta/m_N$  & $M$ (kg) & Size (m) & $|\mathbf{r}_{1}|$ (m) & $|\mathbf{r}_{2}|$ (m) & $|\Delta \nu|$ (Hz) & Ref. \\ \hline
Sun  &  $\frac{0.15}{\Lambda_n}+1.1 \left(\frac{1}{\Lambda_p}+\frac{5\times10^{-4}}{\Lambda_e}\right)  + \frac{8 \times 10^{-4}}{\Lambda_{\gamma }} $ & $2.0 \times 10^{30}$ & $7.0 \times10^{8}$ & $1.47\times10^{11}$ & $1.52\times10^{11}$ & $<0.7$ & \cite{Leefer2013a}\\
Moon  &  $\frac{10}{\Lambda_n}+10 \left(\frac{1}{\Lambda_p}+\frac{5\times10^{-4}}{\Lambda_e}\right)  +\frac{0.03}{\Lambda_{\gamma }}$ & $7.3 \times 10^{22}$ & $1.7\times10^{6}$ & $3.69\times10^{8}$ & $3.99\times10^{8}$  & $<0.6$ & \cite{VanTilburg2015}\\ 
Lead  &   $\frac{126}{\Lambda_n}+82 \left(\frac{1}{\Lambda_p}+\frac{5\times10^{-4}}{\Lambda_e}\right)  + \frac{0.9}{\Lambda _{\gamma }}$ & 300 & $0.38\times0.38\times0.18$ & 0.95 & 1.34 & $<0.3$ & This work\\ 
\end{tabular}
\end{table*}

Constraints on the interaction parameters in Eq.~\eqref{eq4}, as a function of $m_\phi$, are obtained by measuring the Dy transition frequencies in two isotopes, $^{164}$Dy and $^{162}$Dy, in the presence of a massive body at varying distances. 
The differential equation for $\phi$ in a source-free region is $\nabla^2\phi \propto m_\phi^2 \phi$. As a consequence, we cannot generally treat an extended source as a point mass located at its origin, even in a system with spherical symmetry, unless the distance between the apparatus and massive body is always much greater than the dimensions of both. 
We, therefore, define the total scalar field at position $\mathbf{r}$,
\begin{align}
\Phi(\mathbf{r}) &= \int_V n(\mathbf{r'}) \phi(\mathbf{r-r'}) d^3 r' \notag \\ 
  & = -  \frac{\beta}{4\pi}\int_V n(\mathbf{r'}) \frac{e^{-m_\phi |\mathbf{r-r'}|}}{ |\mathbf{r-r'}|} d^3 r' \equiv \beta\,\mathcal{F}( m_\phi, \mathbf{r}), \label{eq7}
\end{align} 
where $n$ is the number density of atoms in the source mass, the integral is over the position vector $\mathbf{r'}$ within the volume of the source mass, and $\beta$ is a source-dependent parameter that is defined in Eq.~\eqref{eq4}. 
We can combine Eqs.~\eqref{eq3}, \eqref{eq6}, and \eqref{eq7} to give
\begin{equation}\label{eq8}
\Delta \nu = K_\alpha \frac{\beta}{\Lambda_\gamma}\left[\mathcal{F}( m_\phi, \mathbf{r}_2) - \mathcal{F}( m_\phi, \mathbf{r}_1)\right].
\end{equation}

In the limit as the scalar-particle mass $m_\phi \rightarrow 0$, the field $\Phi$ has a $1/r$ dependence (for a spherically-symmetric source) that is proportional to the local gravitational potential, $V = - G M /r$, where $G$ is Newton's constant and $M$ is the mass of the source body. 
In this range of scalar-particle masses, the best constraints come from looking for a correlation between $\alpha$ and the varying gravitational potential in the laboratory due to Earth's eccentric orbit about the Sun. 
In Ref.~\cite{Leefer2013a}, it was found that the Dy transition frequency changed by $|\Delta \nu| < 0.7$~Hz as the Earth-Sun distance changed between $1.52\times10^8$~km and $1.47\times10^8$~km. 
In the natural unit system, the average Earth-Sun distance corresponds to a scalar-particle mass of $m_\phi = 1/R \approx 10^{-18}$~eV. 
For $m_\phi \gg 1/R$, the exponential fall-off of $\Phi$ limits the ability of laboratory experiments, which utilise the eccentricity of Earth's orbit around the Sun, to constrain the Yukawa-type interaction parameters.

The changing distance between the Moon and Earth allows one to investigate couplings for larger values of $m_\phi$. 
The Earth-Moon distance is on average $3.84\times10^5$~km, center to center. 
This varies by about 40,000 km with a period of approximately 27.3 days, thus providing an avenue for observing the influence of $\Phi$ on laboratory experiments. 
Complicating the analysis is that the amplitude of this variation is not constant with time, varying between $\sim$30,000 km and $\sim$50,000 km. 
Additionally, the diameter of Earth is 12,700~km. 
Thus the laboratory-Moon distance has a non-negligible daily variation on top of the approximately monthly variation. 
To simplify the analysis, we make use of the observation that Ref.~\cite{VanTilburg2015} constrained variation of the Dy transition frequency for a broad range of oscillation periods, including the daily and monthly lunar periods, at the level of $|\Delta \nu| <0.6$~Hz. 
The monthly variation in the Earth-Moon distance is the most significant, and so we make a conservative bound by using $\sim$30,000 km, which is the minimal seasonal distance variation, as the amplitude of the monthly variation in distance between the Moon and Earth. 

To investigate even larger values of $m_\phi$, we modulated the proximity of a 300 kg lead mass near our experimental setup while measuring the Dy transition frequencies. 
A winch attached to the laboratory ceiling was used to lift and lower the mass next to the apparatus, as shown in Fig.~\ref{fig1}. 
This changed the distance, and hence altered the magnitude of the scalar field. 
In order to minimise impact on the apparatus, contact of the lead test mass with the floor was avoided while taking measurements. 
The weight is anchored at one point on the winch, thus mechanical loads on the building structure do not depend on the position of the lead mass. 
The lead mass is rectangular in shape, with the dimensions L x W x H of 38 cm x 38 cm x 18 cm. 
The center of the mass can be brought to within 95 cm of the atom interaction region by lifting the mass to an equal height with the atoms. 
The mass is alternatively lowered 95 cm towards the floor. 
Using the known experimental and source-mass geometries, the scalar field amplitude $\Phi$ at the position of the Dy atoms was numerically integrated for various values of $m_\phi$ and distance from the Dy atoms using Eq.~\eqref{eq7}. 

During a total time of 80 minutes, the lead was alternated three times back and forth between the up and down positions. 
In each step, the transition frequencies of both Dy isotopes were measured for five minutes per isotope. 
The difference in the frequencies measured at the high and low position was found to be
\begin{align}
\Delta \nu = \begin{cases} \unit[(0.33  \pm 0.20)]{Hz} \quad \text{for $^{162}$Dy,} \\ \unit[(0.03 \pm 0.19 )]{Hz} \quad \text{for $^{164}$Dy.} \end{cases} \notag
\end{align}
Averaging the two measurements and assuming uncorrelated errors gives the result $\Delta \nu = 0.17(14)$~Hz. 
As this result is essentially consistent with zero at $\sim 1\,\sigma$, we conclude that $|\Delta \nu| < 0.3$~Hz. 

\begin{figure*}[t]
\includegraphics[width=8.5cm]{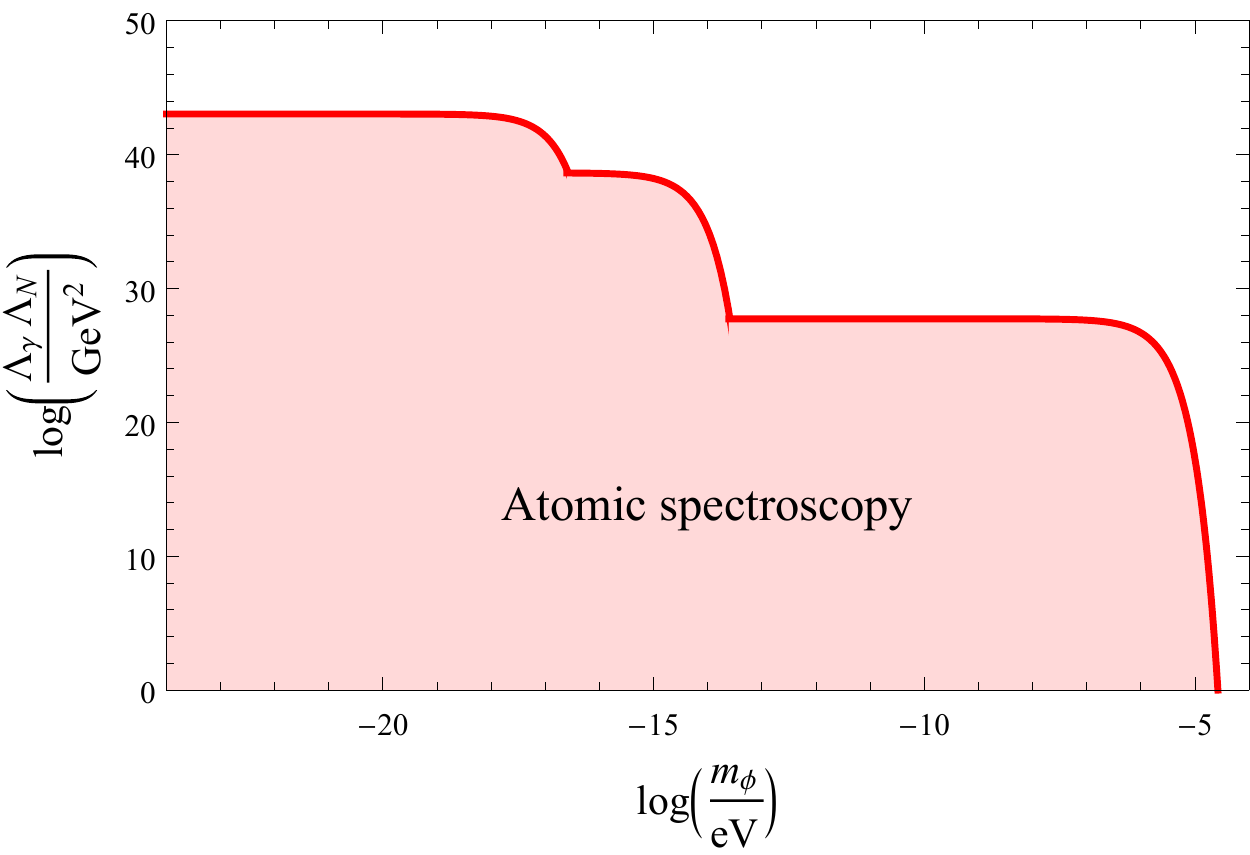}
\includegraphics[width=8.5cm]{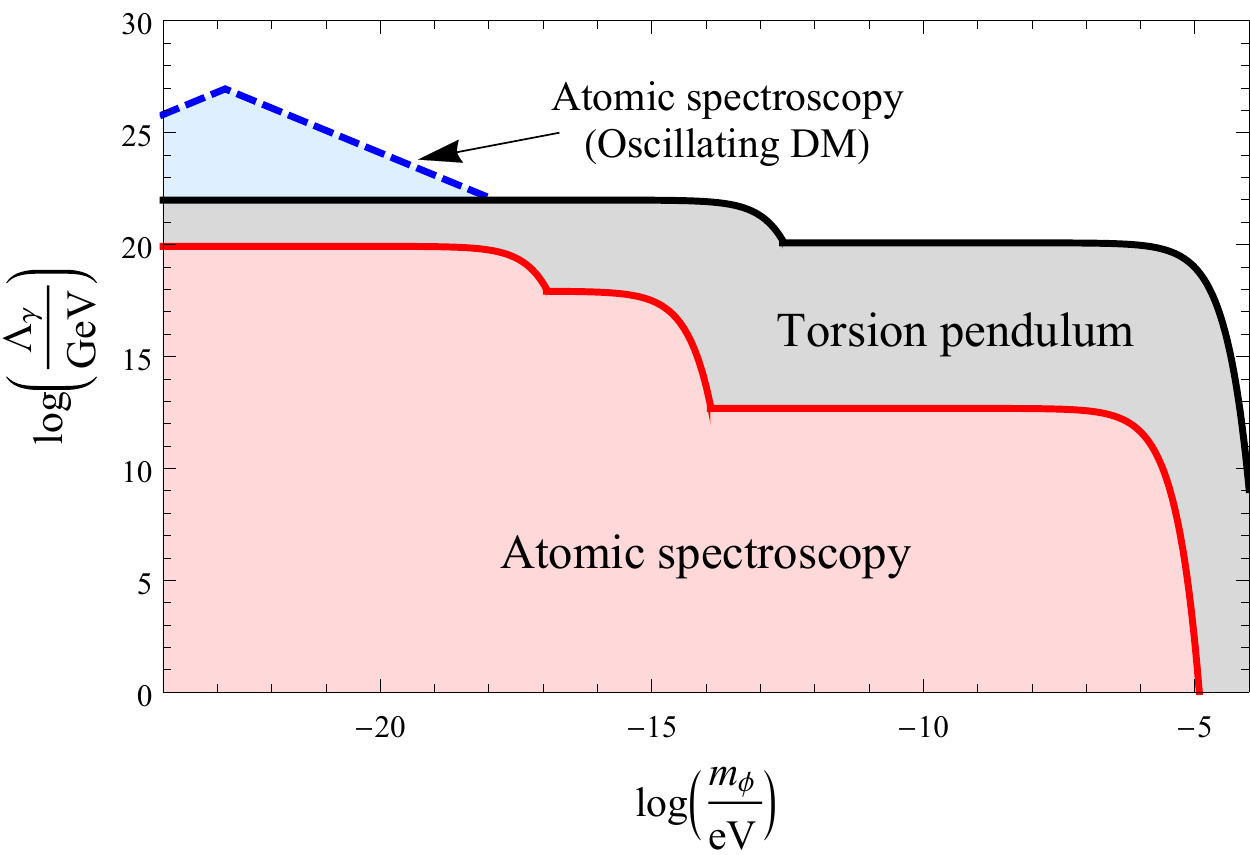}
\includegraphics[width=8.5cm]{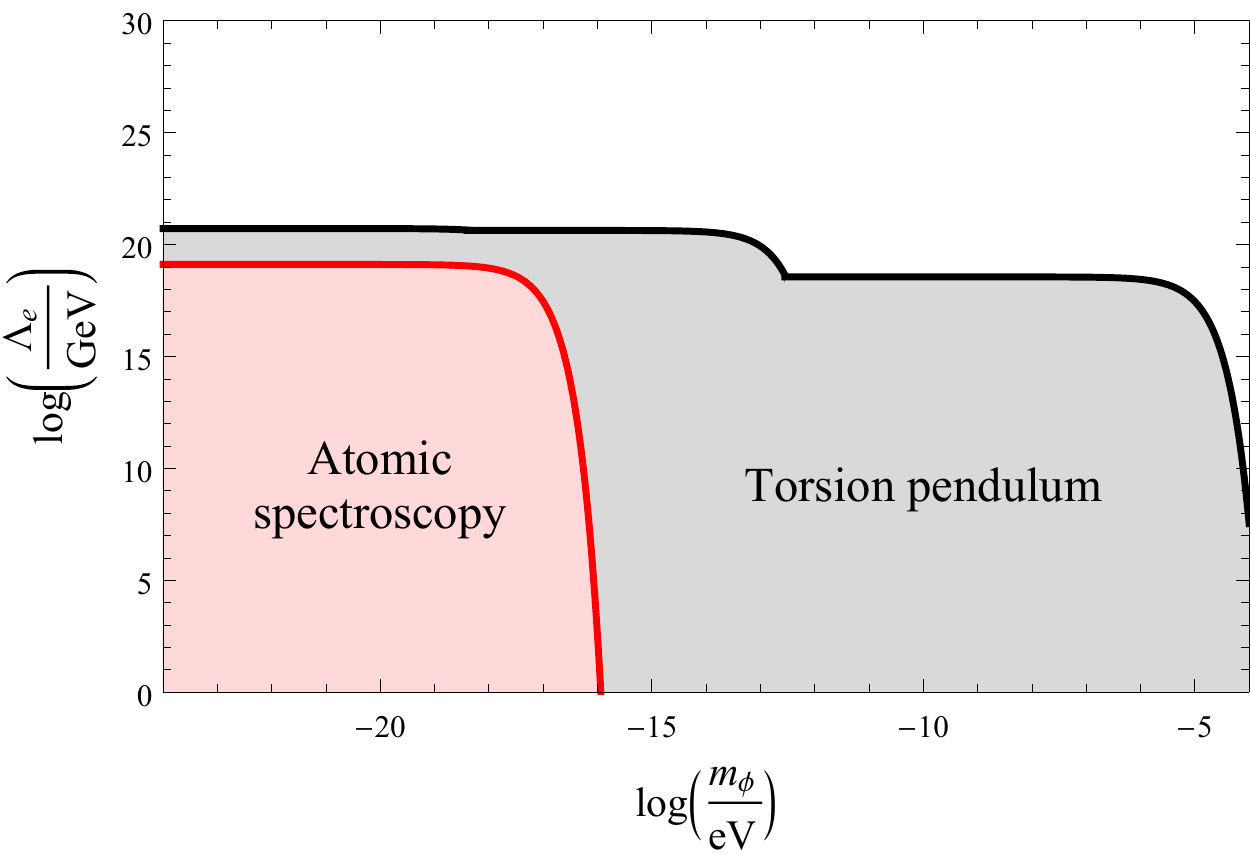}
\includegraphics[width=8.5cm]{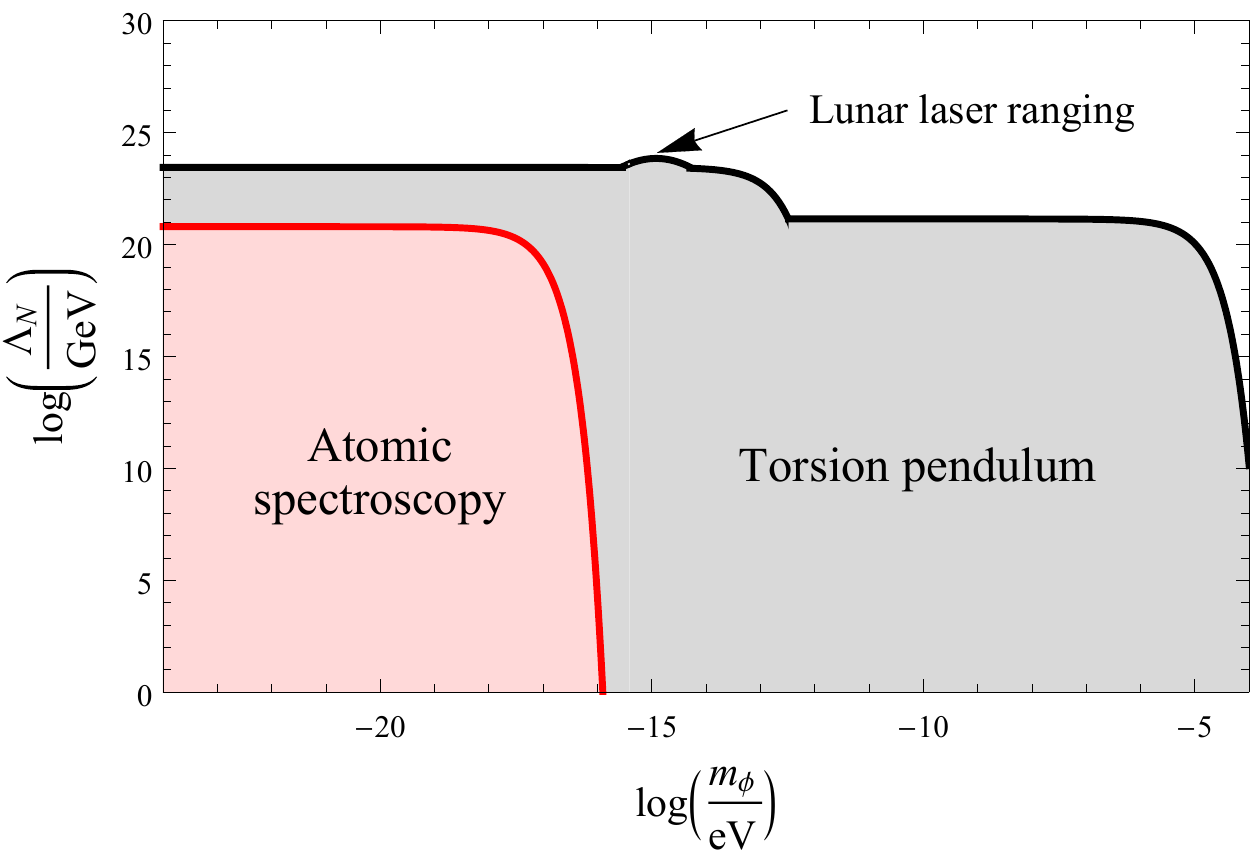}
\caption{\label{fig2}(Color online) 
Limits on the Yukawa-type interactions of the scalar field $\phi$ with the photon, electron and nucleons (assuming an isotopically-invariant interaction), as defined in Eq.~(\ref{eq1}). 
The regions in red correspond to regions of parameters excluded by the present work. 
The regions in grey correspond to existing constraints from searches for anomalous forces due to the exchange of virtual $\phi$ quanta \cite{Moscow1972,Adelberger1999_U238,Schlamminger2008,Adelberger2009,LLR2004}. 
See Table~\ref{table2} for further details. 
A detailed geological and topographical analysis in combination with existing torsion pendulum measurements gives additional constraints (not shown) for $1/R_{\textrm{Earth}} \lesssim m_\phi \lesssim 10^{-7}$~eV (see Refs.~\cite{Schlamminger2008,Adelberger2009,Wagner2012torsion} and the references therein for more details). 
The region in blue corresponds to existing constraints from atomic spectroscopy measurements that search for the effects of a relic coherently oscillating field $\phi = \phi_0 \cos(m_\phi t)$, which saturates the local cold dark matter content \cite{VanTilburg2015,Hees2016DM}. 
}
\end{figure*}

The three limits on $\Delta \nu$ in the presence of varying distances between Dy and the various source bodies (see Table~\ref{table1}) exclude the corresponding combinations of parameters, $\beta / \Lambda_\gamma$, where $\beta$ is a source-dependent function of interaction parameters, defined in Eq.~\eqref{eq4}, as functions of $m_\phi$. 
We present these limits in Table~\ref{table2}. 
We note that the combinations of parameters $\beta / \Lambda_\gamma$ cannot be probed by experiments that search for anomalous forces --- such experiments instead probe the combinations of parameters $(\beta_1/\mu_1 - \beta_2/\mu_2) \beta_3/\mu_3$. 
If, e.g., the source-dependent functions $\beta$ are dominated by the nucleon terms, then Dy spectroscopy measurements probe the combination of parameters $\Lambda_\gamma \Lambda_N$ (see Fig.~\ref{fig2}), while anomalous-force measurements probe the parameter $\Lambda_N^2$, assuming an isotopically-invariant interaction with the nucleons. 

\begin{table*}[t]
\centering
\caption{Summary of limits on the Yukawa-type interactions of the scalar field $\phi$ with the photon, electron and nucleons (assuming an isotopically-invariant interaction), as defined in Eq.~(\ref{eq1}). 
The source-dependent functions $\beta$ for the Sun, the Moon and lead are presented in Table \ref{table1}. 
We have assumed that the elemental composition of Earth is a 1:1:1 ratio of $^{24}$Mg$^{16}$O, $^{28}$Si$^{16}$O$_2$ and $^{56}$Fe by number, and have neglected Earth's internal structure. 
For the assumed elemental compositions of the Sun and the Moon, see the caption to Table \ref{table1}. 
}
\label{table2}
\resizebox{\linewidth}{!}{
\begin{tabular}{ c|c|c|c|c|c|c|c|c }%
Method & System(s)  & Source/Attractor & $\Lambda_\gamma / \beta$ (GeV) & $\Lambda_\gamma$ (GeV) & $\Lambda_e$ (GeV) & $\Lambda_N$ (GeV)  & $m_\phi$ (eV) & Ref. \\ \hline
Atomic spectroscopy  & $^{162,164}$Dy/$^{133}$Cs  & Sun & $1.0 \times10^{43}$  & $8 \times10^{19}$ & --- & ---  & $\lesssim 1 \times 10^{-18}$ & This work \\
Atomic spectroscopy  & $^{1}$H(hyperfine)/$^{133}$Cs,  & Sun & --- & $4 \times10^{19}$ & $1.3 \times10^{19}$ & $6 \times10^{20}$  & $\lesssim 1 \times 10^{-18}$ & This work \\
  & $^{87}$Rb/$^{133}$Cs, $^{87}$Sr/$^{133}$Cs,  &  &  &  &  &  &  &  \\
  & $^{162,164}$Dy/$^{133}$Cs$^a$, $^{199}$Hg$^+$/$^{133}$Cs  &  &  &  &  &  &  &  \\
Atomic spectroscopy  & $^{162,164}$Dy/$^{133}$Cs & Moon & $2 \times10^{37}$  & $8 \times10^{17}$ & --- & ---  & $\lesssim 5 \times 10^{-16}$ & This work \\
Atomic spectroscopy  & $^{162,164}$Dy/$^{133}$Cs  & Lead (300 kg) & $3 \times10^{25}$ & $5 \times10^{12}$ & --- & ---  & $\lesssim 2 \times 10^{-7}$ & This work \\
Torsion pendulum  & Al-Pt, Be-Ti, Be-Al  & Sun &  --- & $5 \times10^{21}$ & $5 \times10^{20}$ & $2 \times10^{23}$  & $\lesssim 1 \times 10^{-18}$ & \cite{Moscow1972,Schlamminger2008,Adelberger2009} \\
Torsion pendulum  & Be-Ti, Be-Al  & Earth & --- & $1.0 \times10^{22}$ & $4 \times10^{20}$ & $3 \times10^{23}$  & $\lesssim 3 \times 10^{-14}$ & \cite{Schlamminger2008,Adelberger2009} \\
Torsion pendulum  & Cu-Pb  & $^{238}$U (3 tonne) & --- & $1.2 \times10^{20}$ & $4 \times10^{18}$ & $1.4 \times10^{21}$  & $\lesssim 2 \times 10^{-6}$ & \cite{Adelberger1999_U238} \\
Lunar laser ranging  & Moon-Earth  & Sun & --- & $3 \times10^{21}$ & $3 \times10^{20}$ & $1.0 \times10^{23}$  & $\lesssim 1 \times 10^{-18}$ & \cite{LLR2004} \\
Lunar laser ranging  & Moon (geodetic precession)  & Earth & --- & $1.2 \times10^{21}$ & $2 \times10^{20}$ & $7 \times10^{23}$  & $\sim 10^{-15}$ & \cite{LLR2004} \\
Atom interferometry  & $^{85}$Rb-$^{87}$Rb  & Earth & --- & $6 \times10^{18}$ & $5 \times10^{17}$ & $5 \times10^{19}$  & $\lesssim 3 \times 10^{-14}$ & \cite{AI2015} \\   
\footnotetext{Only pre-2010 Dy/Cs data are included in this fit.}
\end{tabular}
}
\end{table*}

By assuming in turn that one of the new-physics energy scales in Eq.~(\ref{eq1}) is sufficiently smaller than all of the other energy scales, we can derive constraints on the individual energy scales appearing in Eq.~(\ref{eq1}). 
We present limits on the parameter $\Lambda_\gamma$ from Dy spectroscopy measurements in Fig.~\ref{fig2} and Table~\ref{table2}. 
We have also derived limits on $\Lambda_\gamma$, $\Lambda_e$ and $\Lambda_N$ using data from other atomic spectroscopy measurements, which are described in Ref.~\cite{Guena2012}. 
In Table~\ref{table2}, we summarise all of our derived limits, together with existing limits from anomalous-force searches. 
In the limit as the scalar-particle mass $m_\phi \to 0$, all of our derived limits on $\Lambda_\gamma$, $\Lambda_e$ and $\Lambda_N$ exceed the Planck scale. 
This is significant, since the only other laboratory measurements to achieve super-Planckian sensitivity to the parameters $\Lambda_\gamma$ and $\Lambda_e$ to date have been torsion pendulum experiments.
We find that the most promising individual-parameter constraints from atomic spectroscopy, compared to other methods, are on the parameter $\Lambda_e$; in particular, the constraints on $\Lambda_e$ from atomic spectroscopy are 1.5 orders of magnitude better than constraints from atom interferometry in the limit as $m_\phi \to 0$. 
The reason for this is that the sensitivity of traditional anomalous-force searches 
to $\Lambda_e^2$ is parametrically suppressed relative to the sensitivity to $\Lambda_n^2$ and $\Lambda_p^2$ by the small factor $(m_e/m_N)^2$, while for atomic spectroscopy measurements
, which compare an optical transition frequency with a magnetic hyperfine transition frequency, the parametric suppression is only by the factor $m_e/m_N$.

Following this preliminary work, it is worthwhile to consider which other measurements may provide new valuable information. 
Firstly, one may use different systems in the laboratory. 
Optical frequency measurements in atoms and ions have recently produced frequency references with absolute fractional stability at the $\sim$$10^{-18}$ level after several hours of averaging~\cite{Chou2010,Hinkley2013,Bloom2014,Huntemann2016}. 
This translates into an improvement in sensitivity to variation of $\alpha$, and hence to the parameter $\beta / \Lambda_\gamma$, by approximately 1 order of magnitude with optical transitions in trapped atoms and ions over the results presented here with atomic dysprosium. 
Further improvements may also come from the spectroscopy of highly-charged ions~\cite{Berengut2010,Dzuba2012HCI,Windberger2015}, molecules \cite{Flambaum2007Mol,Ye2008Mol,DeMille2007Mol} and the proposed isomeric transition in $^{229}$Th~\cite{Peik2003,Flambaum2006Th,Kazakov2012}, as well as laser and maser interferometry \cite{Stadnik2015Laser}.

Secondly, one may implement different experimental geometries. 
Shifting the atoms closer to the source mass can also improve sensitivity to larger $m_\phi$. 
By bringing trapped atoms or ions to within 1 mm of the surface of a massive object, one could extend the higher-mass constraints up to $m_\phi \lesssim 10^{-4}$~eV. 
In order to improve sensitivity to lower-mass scalar particles, one should maximise the combination of parameters, $M \Delta \mathcal{F}( m_\phi, \mathbf{r})/N$, where $M$ is the mass of the source body, $N$ is the number of atoms that comprise the source body, and $\mathcal{F}$ is the function defined in Eq.~(\ref{eq7}). 
Measuring the difference in the ratio of two clock frequencies in the laboratory and on GPS satellites (using Earth as the source body) would allow one to probe the scalar-particle masses $m_\phi \lesssim 3 \times 10^{-14}$~eV, while also providing an increase in $M \Delta \mathcal{F} / N$ in the limit as $m_\phi \to 0$ by a factor of $2$ compared with laboratory measurements that look for variations in the ratio of two clock frequencies due to the eccentricity of Earth's orbit around the Sun. 
For measurements that use the Sun as the source body, one may measure the difference in the ratio of two clock frequencies in the laboratory and on a space probe. 
For a probe incident towards the Sun, $M \Delta \mathcal{F} / N$ may be increased by up to 4 orders of magnitude, while for a probe incident away from the Sun, $M \Delta \mathcal{F} / N$ may be increased by up to 1.5 orders of magnitude, compared with laboratory measurements that look for variations in the ratio of two clock frequencies due to the eccentricity of Earth's orbit around the Sun.

An ideal source body should have the highest possible mass density. 
For objects located in the Solar System, the maximum mass density is limited to 22.6 g/cm$^3$ for osmium in the laboratory, which is roughly twice that of lead; 
however, astrophysical objects outside of our Solar System can have much larger mass densities, e.g., white-dwarf stars typically have $\rho \sim 10^6$~g/cm$^3$. 
The comparison of atomic spectra in the vicinity of white-dwarf stars (see, e.g., Ref.~\cite{Berengut2013}) may thus provide more competitive constraints than those derived in the present work for some values of $m_\phi$.

\begin{acknowledgements}
This work is supported by the Australian Research Council (ARC), the DFG Reinhart Koselleck project, and the ERC (Dark-OST Advanced Project). 
N.~L.~was supported by a Marie Curie International Incoming Fellowship within the 7th European Community Framework Programme. 
The authors acknowledge the kind patience and many helpful discussions with Peter Graham and Michael Hohensee in the initial stages of this work. 
N.~L.~acknowledges Holger M\"{u}ller for asking the qualifying exam question that ultimately inspired this work. 
\end{acknowledgements}

\bibliography{scalarDM}

\begin{thebibliography}{55}%
\makeatletter
\providecommand \@ifxundefined [1]{%
 \@ifx{#1\undefined}
}%
\providecommand \@ifnum [1]{%
 \ifnum #1\expandafter \@firstoftwo
 \else \expandafter \@secondoftwo
 \fi
}%
\providecommand \@ifx [1]{%
 \ifx #1\expandafter \@firstoftwo
 \else \expandafter \@secondoftwo
 \fi
}%
\providecommand \natexlab [1]{#1}%
\providecommand \enquote  [1]{``#1''}%
\providecommand \bibnamefont  [1]{#1}%
\providecommand \bibfnamefont [1]{#1}%
\providecommand \citenamefont [1]{#1}%
\providecommand \href@noop [0]{\@secondoftwo}%
\providecommand \href [0]{\begingroup \@sanitize@url \@href}%
\providecommand \@href[1]{\@@startlink{#1}\@@href}%
\providecommand \@@href[1]{\endgroup#1\@@endlink}%
\providecommand \@sanitize@url [0]{\catcode `\\12\catcode `\$12\catcode
  `\&12\catcode `\#12\catcode `\^12\catcode `\_12\catcode `\%12\relax}%
\providecommand \@@startlink[1]{}%
\providecommand \@@endlink[0]{}%
\providecommand \url  [0]{\begingroup\@sanitize@url \@url }%
\providecommand \@url [1]{\endgroup\@href {#1}{\urlprefix }}%
\providecommand \urlprefix  [0]{URL }%
\providecommand \Eprint [0]{\href }%
\providecommand \doibase [0]{http://dx.doi.org/}%
\providecommand \selectlanguage [0]{\@gobble}%
\providecommand \bibinfo  [0]{\@secondoftwo}%
\providecommand \bibfield  [0]{\@secondoftwo}%
\providecommand \translation [1]{[#1]}%
\providecommand \BibitemOpen [0]{}%
\providecommand \bibitemStop [0]{}%
\providecommand \bibitemNoStop [0]{.\EOS\space}%
\providecommand \EOS [0]{\spacefactor3000\relax}%
\providecommand \BibitemShut  [1]{\csname bibitem#1\endcsname}%
\let\auto@bib@innerbib\@empty
\bibitem [{\citenamefont {Hinshaw}\ \emph {et~al.}(2013)\citenamefont {Hinshaw}
  \emph {et~al.}}]{Hinshaw2013}%
  \BibitemOpen
  \bibfield  {author} {\bibinfo {author} {\bibfnamefont {G.}~\bibnamefont
  {Hinshaw}} \emph {et~al.},\ }\href {\doibase 10.1088/0067-0049/208/2/19}
  {\bibfield  {journal} {\bibinfo  {journal} {The Astrophysical Journal
  Supplement Series}\ }\textbf {\bibinfo {volume} {208}},\ \bibinfo {pages}
  {19} (\bibinfo {year} {2013})}\BibitemShut {NoStop}%
\bibitem [{\citenamefont {Ade}\ \emph {et~al.}()\citenamefont {Ade} \emph
  {et~al.}}]{Planck1502Ade}%
  \BibitemOpen
  \bibfield  {author} {\bibinfo {author} {\bibfnamefont {P.}~\bibnamefont
  {Ade}} \emph {et~al.},\ }\href {https://arxiv.org/abs/1502.01589} {\bibinfo
  {journal} {arXiv:1502.01589}\ }\BibitemShut {NoStop}%
\bibitem [{\citenamefont {Essig}\ \emph {et~al.}()\citenamefont {Essig} \emph
  {et~al.}}]{Essig2013}%
  \BibitemOpen
\bibfield  {journal} {  }\bibfield  {author} {\bibinfo {author} {\bibfnamefont
  {R.}~\bibnamefont {Essig}} \emph {et~al.},\ }\href
  {http://arxiv.org/abs/1311.0029} {\emph {\bibinfo {title} {Report of the
  Community Summer Study 2013 (Snowmass) Intensity Frontier New, Light,
  Weakly-Coupled Particles subgroup}}},\ \bibinfo {type} {Tech. Rep.},\ \Eprint
  {http://arxiv.org/abs/1311.0029} {arXiv:1311.0029} \BibitemShut {NoStop}%
\bibitem [{\citenamefont {Ellis}\ \emph {et~al.}(1989)\citenamefont {Ellis},
  \citenamefont {Kalara}, \citenamefont {Olive},\ and\ \citenamefont
  {Wetterich}}]{ELLIS1989}%
  \BibitemOpen
  \bibfield  {author} {\bibinfo {author} {\bibfnamefont {J.}~\bibnamefont
  {Ellis}}, \bibinfo {author} {\bibfnamefont {S.}~\bibnamefont {Kalara}},
  \bibinfo {author} {\bibfnamefont {K.}~\bibnamefont {Olive}}, \ and\ \bibinfo
  {author} {\bibfnamefont {C.}~\bibnamefont {Wetterich}},\ }\href {\doibase
  10.1016/0370-2693(89)90669-2} {\bibfield  {journal} {\bibinfo  {journal}
  {Physics Letters B}\ }\textbf {\bibinfo {volume} {228}},\ \bibinfo {pages}
  {264 } (\bibinfo {year} {1989})}\BibitemShut {NoStop}%
\bibitem [{\citenamefont {Stadnik}\ and\ \citenamefont
  {Flambaum}(2015{\natexlab{a}})}]{Stadnik2015DM-VFCs}%
  \BibitemOpen
  \bibfield  {author} {\bibinfo {author} {\bibfnamefont {Y.~V.}\ \bibnamefont
  {Stadnik}}\ and\ \bibinfo {author} {\bibfnamefont {V.~V.}\ \bibnamefont
  {Flambaum}},\ }\href {\doibase 10.1103/PhysRevLett.115.201301} {\bibfield
  {journal} {\bibinfo  {journal} {Physical Review Letters}\ }\textbf {\bibinfo
  {volume} {115}},\ \bibinfo {pages} {201301} (\bibinfo {year}
  {2015}{\natexlab{a}})}\BibitemShut {NoStop}%
\bibitem [{\citenamefont {Weizs{\"{a}}cker}(1935)}]{Weizsacker1935}%
  \BibitemOpen
  \bibfield  {author} {\bibinfo {author} {\bibfnamefont {C.~F.~V.}\
  \bibnamefont {Weizs{\"{a}}cker}},\ }\href {\doibase 10.1007/BF01337700}
  {\bibfield  {journal} {\bibinfo  {journal} {Zeitschrift f{\"{u}}r Physik}\
  }\textbf {\bibinfo {volume} {96}},\ \bibinfo {pages} {431} (\bibinfo {year}
  {1935})}\BibitemShut {NoStop}%
\bibitem [{\citenamefont {Cottingham}(1963)}]{COTTINGHAM1963}%
  \BibitemOpen
  \bibfield  {author} {\bibinfo {author} {\bibfnamefont {W.}~\bibnamefont
  {Cottingham}},\ }\href {\doibase 10.1016/0003-4916(63)90023-X} {\bibfield
  {journal} {\bibinfo  {journal} {Annals of Physics}\ }\textbf {\bibinfo
  {volume} {25}},\ \bibinfo {pages} {424 } (\bibinfo {year}
  {1963})}\BibitemShut {NoStop}%
\bibitem [{\citenamefont {Gasser}\ and\ \citenamefont
  {Leutwyler}(1982)}]{GASSER1982}%
  \BibitemOpen
  \bibfield  {author} {\bibinfo {author} {\bibfnamefont {J.}~\bibnamefont
  {Gasser}}\ and\ \bibinfo {author} {\bibfnamefont {H.}~\bibnamefont
  {Leutwyler}},\ }\href {\doibase 10.1016/0370-1573(82)90035-7} {\bibfield
  {journal} {\bibinfo  {journal} {Physics Reports}\ }\textbf {\bibinfo {volume}
  {87}},\ \bibinfo {pages} {77 } (\bibinfo {year} {1982})}\BibitemShut
  {NoStop}%
\bibitem [{\citenamefont {Rosenband}\ \emph {et~al.}(2008)\citenamefont
  {Rosenband}, \citenamefont {Hume}, \citenamefont {Schmidt}, \citenamefont
  {Chou}, \citenamefont {Brusch}, \citenamefont {Lorini}, \citenamefont
  {Oskay}, \citenamefont {Drullinger}, \citenamefont {Fortier}, \citenamefont
  {Stalnaker}, \citenamefont {Diddams}, \citenamefont {Swann}, \citenamefont
  {Newbury}, \citenamefont {Itano}, \citenamefont {Wineland},\ and\
  \citenamefont {Bergquist}}]{Rosenband2008}%
  \BibitemOpen
  \bibfield  {author} {\bibinfo {author} {\bibfnamefont {T.}~\bibnamefont
  {Rosenband}}, \bibinfo {author} {\bibfnamefont {D.~B.}\ \bibnamefont {Hume}},
  \bibinfo {author} {\bibfnamefont {P.~O.}\ \bibnamefont {Schmidt}}, \bibinfo
  {author} {\bibfnamefont {C.~W.}\ \bibnamefont {Chou}}, \bibinfo {author}
  {\bibfnamefont {A.}~\bibnamefont {Brusch}}, \bibinfo {author} {\bibfnamefont
  {L.}~\bibnamefont {Lorini}}, \bibinfo {author} {\bibfnamefont {W.~H.}\
  \bibnamefont {Oskay}}, \bibinfo {author} {\bibfnamefont {R.~E.}\ \bibnamefont
  {Drullinger}}, \bibinfo {author} {\bibfnamefont {T.~M.}\ \bibnamefont
  {Fortier}}, \bibinfo {author} {\bibfnamefont {J.~E.}\ \bibnamefont
  {Stalnaker}}, \bibinfo {author} {\bibfnamefont {S.~A.}\ \bibnamefont
  {Diddams}}, \bibinfo {author} {\bibfnamefont {W.~C.}\ \bibnamefont {Swann}},
  \bibinfo {author} {\bibfnamefont {N.~R.}\ \bibnamefont {Newbury}}, \bibinfo
  {author} {\bibfnamefont {W.~M.}\ \bibnamefont {Itano}}, \bibinfo {author}
  {\bibfnamefont {D.~J.}\ \bibnamefont {Wineland}}, \ and\ \bibinfo {author}
  {\bibfnamefont {J.~C.}\ \bibnamefont {Bergquist}},\ }\href {\doibase
  10.1126/science.1154622} {\bibfield  {journal} {\bibinfo  {journal}
  {Science}\ }\textbf {\bibinfo {volume} {319}},\ \bibinfo {pages} {1808}
  (\bibinfo {year} {2008})}\BibitemShut {NoStop}%
\bibitem [{\citenamefont {Webb}\ \emph {et~al.}(2011)\citenamefont {Webb},
  \citenamefont {King}, \citenamefont {Murphy}, \citenamefont {Flambaum},
  \citenamefont {Carswell},\ and\ \citenamefont {Bainbridge}}]{Webb2011}%
  \BibitemOpen
  \bibfield  {author} {\bibinfo {author} {\bibfnamefont {J.~K.}\ \bibnamefont
  {Webb}}, \bibinfo {author} {\bibfnamefont {J.~A.}\ \bibnamefont {King}},
  \bibinfo {author} {\bibfnamefont {M.~T.}\ \bibnamefont {Murphy}}, \bibinfo
  {author} {\bibfnamefont {V.~V.}\ \bibnamefont {Flambaum}}, \bibinfo {author}
  {\bibfnamefont {R.~F.}\ \bibnamefont {Carswell}}, \ and\ \bibinfo {author}
  {\bibfnamefont {M.~B.}\ \bibnamefont {Bainbridge}},\ }\href {\doibase
  10.1103/PhysRevLett.107.191101} {\bibfield  {journal} {\bibinfo  {journal}
  {Physical Review Letters}\ }\textbf {\bibinfo {volume} {107}},\ \bibinfo
  {pages} {191101} (\bibinfo {year} {2011})}\BibitemShut {NoStop}%
\bibitem [{\citenamefont {Gu\'{e}na}\ \emph {et~al.}(2012)\citenamefont
  {Gu\'{e}na}, \citenamefont {Abgrall}, \citenamefont {Rovera}, \citenamefont
  {Rosenbusch}, \citenamefont {Tobar}, \citenamefont {Laurent}, \citenamefont
  {Clairon},\ and\ \citenamefont {Bize}}]{Guena2012}%
  \BibitemOpen
  \bibfield  {author} {\bibinfo {author} {\bibfnamefont {J.}~\bibnamefont
  {Gu\'{e}na}}, \bibinfo {author} {\bibfnamefont {M.}~\bibnamefont {Abgrall}},
  \bibinfo {author} {\bibfnamefont {D.}~\bibnamefont {Rovera}}, \bibinfo
  {author} {\bibfnamefont {P.}~\bibnamefont {Rosenbusch}}, \bibinfo {author}
  {\bibfnamefont {M.~E.}\ \bibnamefont {Tobar}}, \bibinfo {author}
  {\bibfnamefont {P.}~\bibnamefont {Laurent}}, \bibinfo {author} {\bibfnamefont
  {A.}~\bibnamefont {Clairon}}, \ and\ \bibinfo {author} {\bibfnamefont
  {S.}~\bibnamefont {Bize}},\ }\href {\doibase 10.1103/PhysRevLett.109.080801}
  {\bibfield  {journal} {\bibinfo  {journal} {Physical Review Letters}\
  }\textbf {\bibinfo {volume} {109}},\ \bibinfo {pages} {080801} (\bibinfo
  {year} {2012})}\BibitemShut {NoStop}%
\bibitem [{\citenamefont {Leefer}\ \emph {et~al.}(2013)\citenamefont {Leefer},
  \citenamefont {Weber}, \citenamefont {Cing\"{o}z}, \citenamefont
  {Torgerson},\ and\ \citenamefont {Budker}}]{Leefer2013a}%
  \BibitemOpen
  \bibfield  {author} {\bibinfo {author} {\bibfnamefont {N.}~\bibnamefont
  {Leefer}}, \bibinfo {author} {\bibfnamefont {C.~T.~M.}\ \bibnamefont
  {Weber}}, \bibinfo {author} {\bibfnamefont {A.}~\bibnamefont {Cing\"{o}z}},
  \bibinfo {author} {\bibfnamefont {J.~R.}\ \bibnamefont {Torgerson}}, \ and\
  \bibinfo {author} {\bibfnamefont {D.}~\bibnamefont {Budker}},\ }\href
  {\doibase 10.1103/PhysRevLett.111.060801} {\bibfield  {journal} {\bibinfo
  {journal} {Physical Review Letters}\ }\textbf {\bibinfo {volume} {111}},\
  \bibinfo {pages} {060801} (\bibinfo {year} {2013})}\BibitemShut {NoStop}%
\bibitem [{\citenamefont {Berengut}\ \emph {et~al.}(2013)\citenamefont
  {Berengut}, \citenamefont {Flambaum}, \citenamefont {Ong}, \citenamefont
  {Webb}, \citenamefont {Barrow}, \citenamefont {Barstow}, \citenamefont
  {Preval},\ and\ \citenamefont {Holberg}}]{Berengut2013}%
  \BibitemOpen
  \bibfield  {author} {\bibinfo {author} {\bibfnamefont {J.~C.}\ \bibnamefont
  {Berengut}}, \bibinfo {author} {\bibfnamefont {V.~V.}\ \bibnamefont
  {Flambaum}}, \bibinfo {author} {\bibfnamefont {A.}~\bibnamefont {Ong}},
  \bibinfo {author} {\bibfnamefont {J.~K.}\ \bibnamefont {Webb}}, \bibinfo
  {author} {\bibfnamefont {J.~D.}\ \bibnamefont {Barrow}}, \bibinfo {author}
  {\bibfnamefont {M.~A.}\ \bibnamefont {Barstow}}, \bibinfo {author}
  {\bibfnamefont {S.~P.}\ \bibnamefont {Preval}}, \ and\ \bibinfo {author}
  {\bibfnamefont {J.~B.}\ \bibnamefont {Holberg}},\ }\href {\doibase
  10.1103/PhysRevLett.111.010801} {\bibfield  {journal} {\bibinfo  {journal}
  {Physical Review Letters}\ }\textbf {\bibinfo {volume} {111}},\ \bibinfo
  {pages} {010801} (\bibinfo {year} {2013})}\BibitemShut {NoStop}%
\bibitem [{\citenamefont {Huntemann}\ \emph {et~al.}(2014)\citenamefont
  {Huntemann}, \citenamefont {Lipphardt}, \citenamefont {Tamm}, \citenamefont
  {Gerginov}, \citenamefont {Weyers},\ and\ \citenamefont
  {Peik}}]{Huntemann2014}%
  \BibitemOpen
  \bibfield  {author} {\bibinfo {author} {\bibfnamefont {N.}~\bibnamefont
  {Huntemann}}, \bibinfo {author} {\bibfnamefont {B.}~\bibnamefont
  {Lipphardt}}, \bibinfo {author} {\bibfnamefont {C.}~\bibnamefont {Tamm}},
  \bibinfo {author} {\bibfnamefont {V.}~\bibnamefont {Gerginov}}, \bibinfo
  {author} {\bibfnamefont {S.}~\bibnamefont {Weyers}}, \ and\ \bibinfo {author}
  {\bibfnamefont {E.}~\bibnamefont {Peik}},\ }\href {\doibase
  10.1103/PhysRevLett.113.210802} {\bibfield  {journal} {\bibinfo  {journal}
  {Physical Review Letters}\ }\textbf {\bibinfo {volume} {113}},\ \bibinfo
  {pages} {210802} (\bibinfo {year} {2014})}\BibitemShut {NoStop}%
\bibitem [{\citenamefont {Godun}\ \emph {et~al.}(2014)\citenamefont {Godun},
  \citenamefont {Nisbet-Jones}, \citenamefont {Jones}, \citenamefont {King},
  \citenamefont {Johnson}, \citenamefont {Margolis}, \citenamefont {Szymaniec},
  \citenamefont {Lea}, \citenamefont {Bongs},\ and\ \citenamefont
  {Gill}}]{Godun2014}%
  \BibitemOpen
  \bibfield  {author} {\bibinfo {author} {\bibfnamefont {R.}~\bibnamefont
  {Godun}}, \bibinfo {author} {\bibfnamefont {P.}~\bibnamefont {Nisbet-Jones}},
  \bibinfo {author} {\bibfnamefont {J.}~\bibnamefont {Jones}}, \bibinfo
  {author} {\bibfnamefont {S.}~\bibnamefont {King}}, \bibinfo {author}
  {\bibfnamefont {L.}~\bibnamefont {Johnson}}, \bibinfo {author} {\bibfnamefont
  {H.}~\bibnamefont {Margolis}}, \bibinfo {author} {\bibfnamefont
  {K.}~\bibnamefont {Szymaniec}}, \bibinfo {author} {\bibfnamefont
  {S.}~\bibnamefont {Lea}}, \bibinfo {author} {\bibfnamefont {K.}~\bibnamefont
  {Bongs}}, \ and\ \bibinfo {author} {\bibfnamefont {P.}~\bibnamefont {Gill}},\
  }\href {\doibase 10.1103/PhysRevLett.113.210801} {\bibfield  {journal}
  {\bibinfo  {journal} {Physical Review Letters}\ }\textbf {\bibinfo {volume}
  {113}},\ \bibinfo {pages} {210801} (\bibinfo {year} {2014})}\BibitemShut
  {NoStop}%
\bibitem [{\citenamefont {E{\"{o}}tv{\"{o}}s}\ \emph
  {et~al.}(1922)\citenamefont {E{\"{o}}tv{\"{o}}s}, \citenamefont
  {Pek{\'{á}}r},\ and\ \citenamefont {Fekete}}]{Eotvos1922}%
  \BibitemOpen
  \bibfield  {author} {\bibinfo {author} {\bibfnamefont {R.}~\bibnamefont
  {E{\"{o}}tv{\"{o}}s}}, \bibinfo {author} {\bibfnamefont {D.}~\bibnamefont
  {Pek{\'{á}}r}}, \ and\ \bibinfo {author} {\bibfnamefont {E.}~\bibnamefont
  {Fekete}},\ }\href@noop {} {\bibfield  {journal} {\bibinfo  {journal} {Annals
  of Physics}\ }\textbf {\bibinfo {volume} {373}},\ \bibinfo {pages} {11}
  (\bibinfo {year} {1922})}\BibitemShut {NoStop}%
\bibitem [{\citenamefont {Roll}\ \emph {et~al.}(1964)\citenamefont {Roll},
  \citenamefont {Krotkov},\ and\ \citenamefont {Dicke}}]{Princeton1964}%
  \BibitemOpen
  \bibfield  {author} {\bibinfo {author} {\bibfnamefont {P.}~\bibnamefont
  {Roll}}, \bibinfo {author} {\bibfnamefont {R.}~\bibnamefont {Krotkov}}, \
  and\ \bibinfo {author} {\bibfnamefont {R.}~\bibnamefont {Dicke}},\
  }\href@noop {} {\bibfield  {journal} {\bibinfo  {journal} {Annals of
  Physics}\ }\textbf {\bibinfo {volume} {26}},\ \bibinfo {pages} {442}
  (\bibinfo {year} {1964})}\BibitemShut {NoStop}%
\bibitem [{\citenamefont {Braginsky}\ and\ \citenamefont
  {Panov}(1972)}]{Moscow1972}%
  \BibitemOpen
  \bibfield  {author} {\bibinfo {author} {\bibfnamefont {V.}~\bibnamefont
  {Braginsky}}\ and\ \bibinfo {author} {\bibfnamefont {V.}~\bibnamefont
  {Panov}},\ }\href@noop {} {\bibfield  {journal} {\bibinfo  {journal} {JETP}\
  }\textbf {\bibinfo {volume} {34}},\ \bibinfo {pages} {463} (\bibinfo {year}
  {1972})}\BibitemShut {NoStop}%
\bibitem [{\citenamefont {Smith}\ \emph {et~al.}(1999)\citenamefont {Smith},
  \citenamefont {Hoyle}, \citenamefont {Gundlach}, \citenamefont {Adelberger},
  \citenamefont {Heckel},\ and\ \citenamefont {Swanson}}]{Adelberger1999_U238}%
  \BibitemOpen
  \bibfield  {author} {\bibinfo {author} {\bibfnamefont {G.~L.}\ \bibnamefont
  {Smith}}, \bibinfo {author} {\bibfnamefont {C.~D.}\ \bibnamefont {Hoyle}},
  \bibinfo {author} {\bibfnamefont {J.~H.}\ \bibnamefont {Gundlach}}, \bibinfo
  {author} {\bibfnamefont {E.~G.}\ \bibnamefont {Adelberger}}, \bibinfo
  {author} {\bibfnamefont {B.~R.}\ \bibnamefont {Heckel}}, \ and\ \bibinfo
  {author} {\bibfnamefont {H.~E.}\ \bibnamefont {Swanson}},\ }\href {\doibase
  10.1103/PhysRevD.61.022001} {\bibfield  {journal} {\bibinfo  {journal}
  {Physical Review D}\ }\textbf {\bibinfo {volume} {61}},\ \bibinfo {pages}
  {022001} (\bibinfo {year} {1999})}\BibitemShut {NoStop}%
\bibitem [{\citenamefont {Schlamminger}\ \emph {et~al.}(2008)\citenamefont
  {Schlamminger}, \citenamefont {Choi}, \citenamefont {Wagner}, \citenamefont
  {Gundlach},\ and\ \citenamefont {Adelberger}}]{Schlamminger2008}%
  \BibitemOpen
  \bibfield  {author} {\bibinfo {author} {\bibfnamefont {S.}~\bibnamefont
  {Schlamminger}}, \bibinfo {author} {\bibfnamefont {K.-Y.}\ \bibnamefont
  {Choi}}, \bibinfo {author} {\bibfnamefont {T.}~\bibnamefont {Wagner}},
  \bibinfo {author} {\bibfnamefont {J.}~\bibnamefont {Gundlach}}, \ and\
  \bibinfo {author} {\bibfnamefont {E.}~\bibnamefont {Adelberger}},\ }\href
  {\doibase 10.1103/PhysRevLett.100.041101} {\bibfield  {journal} {\bibinfo
  {journal} {Physical Review Letters}\ }\textbf {\bibinfo {volume} {100}},\
  \bibinfo {pages} {041101} (\bibinfo {year} {2008})}\BibitemShut {NoStop}%
\bibitem [{\citenamefont {Adelberger}\ \emph {et~al.}(2009)\citenamefont
  {Adelberger}, \citenamefont {Gundlach}, \citenamefont {Heckel}, \citenamefont
  {Hoedl},\ and\ \citenamefont {Schlamminger}}]{Adelberger2009}%
  \BibitemOpen
  \bibfield  {author} {\bibinfo {author} {\bibfnamefont {E.}~\bibnamefont
  {Adelberger}}, \bibinfo {author} {\bibfnamefont {J.}~\bibnamefont
  {Gundlach}}, \bibinfo {author} {\bibfnamefont {B.}~\bibnamefont {Heckel}},
  \bibinfo {author} {\bibfnamefont {S.}~\bibnamefont {Hoedl}}, \ and\ \bibinfo
  {author} {\bibfnamefont {S.}~\bibnamefont {Schlamminger}},\ }\href {\doibase
  10.1016/j.ppnp.2008.08.002} {\bibfield  {journal} {\bibinfo  {journal}
  {Progress in Particle and Nuclear Physics}\ }\textbf {\bibinfo {volume}
  {62}},\ \bibinfo {pages} {102} (\bibinfo {year} {2009})}\BibitemShut
  {NoStop}%
\bibitem [{\citenamefont {Wagner}\ \emph {et~al.}(2012)\citenamefont {Wagner},
  \citenamefont {Schlamminger}, \citenamefont {Gundlach},\ and\ \citenamefont
  {Adelberger}}]{Wagner2012torsion}%
  \BibitemOpen
  \bibfield  {author} {\bibinfo {author} {\bibfnamefont {T.~A.}\ \bibnamefont
  {Wagner}}, \bibinfo {author} {\bibfnamefont {S.}~\bibnamefont
  {Schlamminger}}, \bibinfo {author} {\bibfnamefont {J.}~\bibnamefont
  {Gundlach}}, \ and\ \bibinfo {author} {\bibfnamefont {E.}~\bibnamefont
  {Adelberger}},\ }\href {http://stacks.iop.org/0264-9381/29/i=18/a=184002}
  {\bibfield  {journal} {\bibinfo  {journal} {Classical and Quantum Gravity}\
  }\textbf {\bibinfo {volume} {29}},\ \bibinfo {pages} {184002} (\bibinfo
  {year} {2012})}\BibitemShut {NoStop}%
\bibitem [{\citenamefont {Talmadge}\ \emph {et~al.}(1988)\citenamefont
  {Talmadge}, \citenamefont {Berthias}, \citenamefont {Hellings},\ and\
  \citenamefont {Standish}}]{LLR1988}%
  \BibitemOpen
  \bibfield  {author} {\bibinfo {author} {\bibfnamefont {C.}~\bibnamefont
  {Talmadge}}, \bibinfo {author} {\bibfnamefont {J.~P.}\ \bibnamefont
  {Berthias}}, \bibinfo {author} {\bibfnamefont {R.~W.}\ \bibnamefont
  {Hellings}}, \ and\ \bibinfo {author} {\bibfnamefont {E.~M.}\ \bibnamefont
  {Standish}},\ }\href {\doibase 10.1103/PhysRevLett.61.1159} {\bibfield
  {journal} {\bibinfo  {journal} {Physical Review Letters}\ }\textbf {\bibinfo
  {volume} {61}},\ \bibinfo {pages} {1159} (\bibinfo {year}
  {1988})}\BibitemShut {NoStop}%
\bibitem [{\citenamefont {Williams}\ \emph {et~al.}(1996)\citenamefont
  {Williams}, \citenamefont {Newhall},\ and\ \citenamefont {Dickey}}]{LLR1996}%
  \BibitemOpen
  \bibfield  {author} {\bibinfo {author} {\bibfnamefont {J.~G.}\ \bibnamefont
  {Williams}}, \bibinfo {author} {\bibfnamefont {X.~X.}\ \bibnamefont
  {Newhall}}, \ and\ \bibinfo {author} {\bibfnamefont {J.~O.}\ \bibnamefont
  {Dickey}},\ }\href {\doibase 10.1103/PhysRevD.53.6730} {\bibfield  {journal}
  {\bibinfo  {journal} {Physical Review D}\ }\textbf {\bibinfo {volume} {53}},\
  \bibinfo {pages} {6730} (\bibinfo {year} {1996})}\BibitemShut {NoStop}%
\bibitem [{\citenamefont {Williams}\ \emph {et~al.}(2004)\citenamefont
  {Williams}, \citenamefont {Turyshev},\ and\ \citenamefont {Boggs}}]{LLR2004}%
  \BibitemOpen
  \bibfield  {author} {\bibinfo {author} {\bibfnamefont {J.~G.}\ \bibnamefont
  {Williams}}, \bibinfo {author} {\bibfnamefont {S.~G.}\ \bibnamefont
  {Turyshev}}, \ and\ \bibinfo {author} {\bibfnamefont {D.~H.}\ \bibnamefont
  {Boggs}},\ }\href {\doibase 10.1103/PhysRevLett.93.261101} {\bibfield
  {journal} {\bibinfo  {journal} {Physical Review Letters}\ }\textbf {\bibinfo
  {volume} {93}},\ \bibinfo {pages} {261101} (\bibinfo {year}
  {2004})}\BibitemShut {NoStop}%
\bibitem [{\citenamefont {Fray}\ \emph {et~al.}(2004)\citenamefont {Fray},
  \citenamefont {Diez}, \citenamefont {H\"ansch},\ and\ \citenamefont
  {Weitz}}]{AI2004}%
  \BibitemOpen
  \bibfield  {author} {\bibinfo {author} {\bibfnamefont {S.}~\bibnamefont
  {Fray}}, \bibinfo {author} {\bibfnamefont {C.~A.}\ \bibnamefont {Diez}},
  \bibinfo {author} {\bibfnamefont {T.~W.}\ \bibnamefont {H\"ansch}}, \ and\
  \bibinfo {author} {\bibfnamefont {M.}~\bibnamefont {Weitz}},\ }\href
  {\doibase 10.1103/PhysRevLett.93.240404} {\bibfield  {journal} {\bibinfo
  {journal} {Physical Review Letters}\ }\textbf {\bibinfo {volume} {93}},\
  \bibinfo {pages} {240404} (\bibinfo {year} {2004})}\BibitemShut {NoStop}%
\bibitem [{\citenamefont {Hohensee}\ \emph
  {et~al.}(2013{\natexlab{a}})\citenamefont {Hohensee}, \citenamefont
  {M\"uller},\ and\ \citenamefont {Wiringa}}]{Hohensee2013AI}%
  \BibitemOpen
  \bibfield  {author} {\bibinfo {author} {\bibfnamefont {M.~A.}\ \bibnamefont
  {Hohensee}}, \bibinfo {author} {\bibfnamefont {H.}~\bibnamefont {M\"uller}},
  \ and\ \bibinfo {author} {\bibfnamefont {R.~B.}\ \bibnamefont {Wiringa}},\
  }\href {\doibase 10.1103/PhysRevLett.111.151102} {\bibfield  {journal}
  {\bibinfo  {journal} {Physical Review Letters}\ }\textbf {\bibinfo {volume}
  {111}},\ \bibinfo {pages} {151102} (\bibinfo {year}
  {2013}{\natexlab{a}})}\BibitemShut {NoStop}%
\bibitem [{\citenamefont {Bonnin}\ \emph {et~al.}(2013)\citenamefont {Bonnin},
  \citenamefont {Zahzam}, \citenamefont {Bidel},\ and\ \citenamefont
  {Bresson}}]{AI2013}%
  \BibitemOpen
  \bibfield  {author} {\bibinfo {author} {\bibfnamefont {A.}~\bibnamefont
  {Bonnin}}, \bibinfo {author} {\bibfnamefont {N.}~\bibnamefont {Zahzam}},
  \bibinfo {author} {\bibfnamefont {Y.}~\bibnamefont {Bidel}}, \ and\ \bibinfo
  {author} {\bibfnamefont {A.}~\bibnamefont {Bresson}},\ }\href {\doibase
  10.1103/PhysRevA.88.043615} {\bibfield  {journal} {\bibinfo  {journal}
  {Physical Review A}\ }\textbf {\bibinfo {volume} {88}},\ \bibinfo {pages}
  {043615} (\bibinfo {year} {2013})}\BibitemShut {NoStop}%
\bibitem [{\citenamefont {Schlippert}\ \emph {et~al.}(2014)\citenamefont
  {Schlippert}, \citenamefont {Hartwig}, \citenamefont {Albers}, \citenamefont
  {Richardson}, \citenamefont {Schubert}, \citenamefont {Roura}, \citenamefont
  {Schleich}, \citenamefont {Ertmer},\ and\ \citenamefont
  {Rasel}}]{AI2014Rasel}%
  \BibitemOpen
  \bibfield  {author} {\bibinfo {author} {\bibfnamefont {D.}~\bibnamefont
  {Schlippert}}, \bibinfo {author} {\bibfnamefont {J.}~\bibnamefont {Hartwig}},
  \bibinfo {author} {\bibfnamefont {H.}~\bibnamefont {Albers}}, \bibinfo
  {author} {\bibfnamefont {L.~L.}\ \bibnamefont {Richardson}}, \bibinfo
  {author} {\bibfnamefont {C.}~\bibnamefont {Schubert}}, \bibinfo {author}
  {\bibfnamefont {A.}~\bibnamefont {Roura}}, \bibinfo {author} {\bibfnamefont
  {W.~P.}\ \bibnamefont {Schleich}}, \bibinfo {author} {\bibfnamefont
  {W.}~\bibnamefont {Ertmer}}, \ and\ \bibinfo {author} {\bibfnamefont {E.~M.}\
  \bibnamefont {Rasel}},\ }\href {\doibase 10.1103/PhysRevLett.112.203002}
  {\bibfield  {journal} {\bibinfo  {journal} {Physical Review Letters}\
  }\textbf {\bibinfo {volume} {112}},\ \bibinfo {pages} {203002} (\bibinfo
  {year} {2014})}\BibitemShut {NoStop}%
\bibitem [{\citenamefont {Tarallo}\ \emph {et~al.}(2014)\citenamefont
  {Tarallo}, \citenamefont {Mazzoni}, \citenamefont {Poli}, \citenamefont
  {Sutyrin}, \citenamefont {Zhang},\ and\ \citenamefont {Tino}}]{AI2014Tino}%
  \BibitemOpen
  \bibfield  {author} {\bibinfo {author} {\bibfnamefont {M.~G.}\ \bibnamefont
  {Tarallo}}, \bibinfo {author} {\bibfnamefont {T.}~\bibnamefont {Mazzoni}},
  \bibinfo {author} {\bibfnamefont {N.}~\bibnamefont {Poli}}, \bibinfo {author}
  {\bibfnamefont {D.~V.}\ \bibnamefont {Sutyrin}}, \bibinfo {author}
  {\bibfnamefont {X.}~\bibnamefont {Zhang}}, \ and\ \bibinfo {author}
  {\bibfnamefont {G.~M.}\ \bibnamefont {Tino}},\ }\href {\doibase
  10.1103/PhysRevLett.113.023005} {\bibfield  {journal} {\bibinfo  {journal}
  {Physical Review Letters}\ }\textbf {\bibinfo {volume} {113}},\ \bibinfo
  {pages} {023005} (\bibinfo {year} {2014})}\BibitemShut {NoStop}%
\bibitem [{\citenamefont {Tino}\ and\ \citenamefont
  {Kasevich}(2014)}]{Tino2014atom}%
  \BibitemOpen
  \bibfield  {author} {\bibinfo {author} {\bibfnamefont {G.}~\bibnamefont
  {Tino}}\ and\ \bibinfo {author} {\bibfnamefont {M.}~\bibnamefont
  {Kasevich}},\ }\href {https://books.google.com.au/books?id=OV7IrQEACAAJ}
  {\emph {\bibinfo {title} {Atom Interferometry}}},\ Proceedings of the
  International School of Physics "Enrico Fermi"\ (\bibinfo  {publisher} {Ios
  PressInc},\ \bibinfo {year} {2014})\BibitemShut {NoStop}%
\bibitem [{\citenamefont {Zhou}\ \emph {et~al.}(2015)\citenamefont {Zhou},
  \citenamefont {Long}, \citenamefont {Tang}, \citenamefont {Chen},
  \citenamefont {Gao}, \citenamefont {Peng}, \citenamefont {Duan},
  \citenamefont {Zhong}, \citenamefont {Xiong}, \citenamefont {Wang},
  \citenamefont {Zhang},\ and\ \citenamefont {Zhan}}]{AI2015}%
  \BibitemOpen
  \bibfield  {author} {\bibinfo {author} {\bibfnamefont {L.}~\bibnamefont
  {Zhou}}, \bibinfo {author} {\bibfnamefont {S.}~\bibnamefont {Long}}, \bibinfo
  {author} {\bibfnamefont {B.}~\bibnamefont {Tang}}, \bibinfo {author}
  {\bibfnamefont {X.}~\bibnamefont {Chen}}, \bibinfo {author} {\bibfnamefont
  {F.}~\bibnamefont {Gao}}, \bibinfo {author} {\bibfnamefont {W.}~\bibnamefont
  {Peng}}, \bibinfo {author} {\bibfnamefont {W.}~\bibnamefont {Duan}}, \bibinfo
  {author} {\bibfnamefont {J.}~\bibnamefont {Zhong}}, \bibinfo {author}
  {\bibfnamefont {Z.}~\bibnamefont {Xiong}}, \bibinfo {author} {\bibfnamefont
  {J.}~\bibnamefont {Wang}}, \bibinfo {author} {\bibfnamefont {Y.}~\bibnamefont
  {Zhang}}, \ and\ \bibinfo {author} {\bibfnamefont {M.}~\bibnamefont {Zhan}},\
  }\href {\doibase 10.1103/PhysRevLett.115.013004} {\bibfield  {journal}
  {\bibinfo  {journal} {Physical Review Letters}\ }\textbf {\bibinfo {volume}
  {115}},\ \bibinfo {pages} {013004} (\bibinfo {year} {2015})}\BibitemShut
  {NoStop}%
\bibitem [{\citenamefont {Hamilton}\ \emph {et~al.}(2015)\citenamefont
  {Hamilton}, \citenamefont {Jaffe}, \citenamefont {Haslinger}, \citenamefont
  {Simmons}, \citenamefont {M{\"u}ller},\ and\ \citenamefont
  {Khoury}}]{Hamilton2015AI}%
  \BibitemOpen
  \bibfield  {author} {\bibinfo {author} {\bibfnamefont {P.}~\bibnamefont
  {Hamilton}}, \bibinfo {author} {\bibfnamefont {M.}~\bibnamefont {Jaffe}},
  \bibinfo {author} {\bibfnamefont {P.}~\bibnamefont {Haslinger}}, \bibinfo
  {author} {\bibfnamefont {Q.}~\bibnamefont {Simmons}}, \bibinfo {author}
  {\bibfnamefont {H.}~\bibnamefont {M{\"u}ller}}, \ and\ \bibinfo {author}
  {\bibfnamefont {J.}~\bibnamefont {Khoury}},\ }\href {\doibase
  10.1126/science.aaa8883} {\bibfield  {journal} {\bibinfo  {journal}
  {Science}\ }\textbf {\bibinfo {volume} {349}},\ \bibinfo {pages} {849}
  (\bibinfo {year} {2015})}\BibitemShut {NoStop}%
\bibitem [{\citenamefont {Budker}\ \emph {et~al.}(1994)\citenamefont {Budker},
  \citenamefont {DeMille}, \citenamefont {Commins},\ and\ \citenamefont
  {Zolotorev}}]{Budker1994}%
  \BibitemOpen
  \bibfield  {author} {\bibinfo {author} {\bibfnamefont {D.}~\bibnamefont
  {Budker}}, \bibinfo {author} {\bibfnamefont {D.}~\bibnamefont {DeMille}},
  \bibinfo {author} {\bibfnamefont {E.}~\bibnamefont {Commins}}, \ and\
  \bibinfo {author} {\bibfnamefont {M.}~\bibnamefont {Zolotorev}},\ }\href
  {\doibase 10.1103/PhysRevA.50.132} {\bibfield  {journal} {\bibinfo  {journal}
  {Physical Review A}\ }\textbf {\bibinfo {volume} {50}},\ \bibinfo {pages}
  {132} (\bibinfo {year} {1994})}\BibitemShut {NoStop}%
\bibitem [{\citenamefont {Hohensee}\ \emph
  {et~al.}(2013{\natexlab{b}})\citenamefont {Hohensee}, \citenamefont {Leefer},
  \citenamefont {Budker}, \citenamefont {Harabati}, \citenamefont {Dzuba},\
  and\ \citenamefont {Flambaum}}]{Hohensee2013}%
  \BibitemOpen
  \bibfield  {author} {\bibinfo {author} {\bibfnamefont {M.~A.}\ \bibnamefont
  {Hohensee}}, \bibinfo {author} {\bibfnamefont {N.}~\bibnamefont {Leefer}},
  \bibinfo {author} {\bibfnamefont {D.}~\bibnamefont {Budker}}, \bibinfo
  {author} {\bibfnamefont {C.}~\bibnamefont {Harabati}}, \bibinfo {author}
  {\bibfnamefont {V.~A.}\ \bibnamefont {Dzuba}}, \ and\ \bibinfo {author}
  {\bibfnamefont {V.~V.}\ \bibnamefont {Flambaum}},\ }\href {\doibase
  10.1103/PhysRevLett.111.050401} {\bibfield  {journal} {\bibinfo  {journal}
  {Physical Review Letters}\ }\textbf {\bibinfo {volume} {111}},\ \bibinfo
  {pages} {050401} (\bibinfo {year} {2013}{\natexlab{b}})}\BibitemShut
  {NoStop}%
\bibitem [{\citenamefont {Van~Tilburg}\ \emph {et~al.}(2015)\citenamefont
  {Van~Tilburg}, \citenamefont {Leefer}, \citenamefont {Bougas},\ and\
  \citenamefont {Budker}}]{VanTilburg2015}%
  \BibitemOpen
  \bibfield  {author} {\bibinfo {author} {\bibfnamefont {K.}~\bibnamefont
  {Van~Tilburg}}, \bibinfo {author} {\bibfnamefont {N.}~\bibnamefont {Leefer}},
  \bibinfo {author} {\bibfnamefont {L.}~\bibnamefont {Bougas}}, \ and\ \bibinfo
  {author} {\bibfnamefont {D.}~\bibnamefont {Budker}},\ }\href {\doibase
  10.1103/PhysRevLett.115.011802} {\bibfield  {journal} {\bibinfo  {journal}
  {Physical Review Letters}\ }\textbf {\bibinfo {volume} {115}},\ \bibinfo
  {pages} {011802} (\bibinfo {year} {2015})}\BibitemShut {NoStop}%
\bibitem [{\citenamefont {Dzuba}\ \emph
  {et~al.}(1999{\natexlab{a}})\citenamefont {Dzuba}, \citenamefont {Flambaum},\
  and\ \citenamefont {Webb}}]{Dzuba1999a}%
  \BibitemOpen
  \bibfield  {author} {\bibinfo {author} {\bibfnamefont {V.~A.}\ \bibnamefont
  {Dzuba}}, \bibinfo {author} {\bibfnamefont {V.~V.}\ \bibnamefont {Flambaum}},
  \ and\ \bibinfo {author} {\bibfnamefont {J.~K.}\ \bibnamefont {Webb}},\
  }\href {\doibase 10.1103/PhysRevLett.82.888} {\bibfield  {journal} {\bibinfo
  {journal} {Physical Review Letters}\ }\textbf {\bibinfo {volume} {82}},\
  \bibinfo {pages} {888} (\bibinfo {year} {1999}{\natexlab{a}})}\BibitemShut
  {NoStop}%
\bibitem [{\citenamefont {Dzuba}\ \emph
  {et~al.}(1999{\natexlab{b}})\citenamefont {Dzuba}, \citenamefont {Flambaum},\
  and\ \citenamefont {Webb}}]{Dzuba1999}%
  \BibitemOpen
  \bibfield  {author} {\bibinfo {author} {\bibfnamefont {V.~A.}\ \bibnamefont
  {Dzuba}}, \bibinfo {author} {\bibfnamefont {V.~V.}\ \bibnamefont {Flambaum}},
  \ and\ \bibinfo {author} {\bibfnamefont {J.~K.}\ \bibnamefont {Webb}},\
  }\href {\doibase 10.1103/PhysRevA.59.230} {\bibfield  {journal} {\bibinfo
  {journal} {Physical Review A}\ }\textbf {\bibinfo {volume} {59}},\ \bibinfo
  {pages} {230} (\bibinfo {year} {1999}{\natexlab{b}})}\BibitemShut {NoStop}%
\bibitem [{\citenamefont {Dzuba}\ \emph {et~al.}(2003)\citenamefont {Dzuba},
  \citenamefont {Flambaum},\ and\ \citenamefont {Marchenko}}]{Dzuba2003}%
  \BibitemOpen
  \bibfield  {author} {\bibinfo {author} {\bibfnamefont {V.~A.}\ \bibnamefont
  {Dzuba}}, \bibinfo {author} {\bibfnamefont {V.~V.}\ \bibnamefont {Flambaum}},
  \ and\ \bibinfo {author} {\bibfnamefont {M.~V.}\ \bibnamefont {Marchenko}},\
  }\href {\doibase 10.1103/PhysRevA.68.022506} {\bibfield  {journal} {\bibinfo
  {journal} {Physical Review A}\ }\textbf {\bibinfo {volume} {68}},\ \bibinfo
  {pages} {022506} (\bibinfo {year} {2003})}\BibitemShut {NoStop}%
\bibitem [{\citenamefont {Dzuba}\ and\ \citenamefont
  {Flambaum}(2008)}]{Dzuba2008Dy}%
  \BibitemOpen
  \bibfield  {author} {\bibinfo {author} {\bibfnamefont {V.~A.}\ \bibnamefont
  {Dzuba}}\ and\ \bibinfo {author} {\bibfnamefont {V.~V.}\ \bibnamefont
  {Flambaum}},\ }\href {\doibase 10.1103/PhysRevA.77.012515} {\bibfield
  {journal} {\bibinfo  {journal} {Physical Review A}\ }\textbf {\bibinfo
  {volume} {77}},\ \bibinfo {pages} {012515} (\bibinfo {year}
  {2008})}\BibitemShut {NoStop}%
\bibitem [{\citenamefont {Hees}\ \emph {et~al.}()\citenamefont {Hees},
  \citenamefont {Gu{\'e}na}, \citenamefont {Abgrall}, \citenamefont {Bize},\
  and\ \citenamefont {Wolf}}]{Hees2016DM}%
  \BibitemOpen
  \bibfield  {author} {\bibinfo {author} {\bibfnamefont {A.}~\bibnamefont
  {Hees}}, \bibinfo {author} {\bibfnamefont {J.}~\bibnamefont {Gu{\'e}na}},
  \bibinfo {author} {\bibfnamefont {M.}~\bibnamefont {Abgrall}}, \bibinfo
  {author} {\bibfnamefont {S.}~\bibnamefont {Bize}}, \ and\ \bibinfo {author}
  {\bibfnamefont {P.}~\bibnamefont {Wolf}},\ }\href
  {https://arxiv.org/abs/1604.08514} {\bibinfo  {journal} {arXiv:1604.08514}\
  }\BibitemShut {NoStop}%
\bibitem [{\citenamefont {Chou}\ \emph {et~al.}(2010)\citenamefont {Chou},
  \citenamefont {Hume}, \citenamefont {Koelemeij}, \citenamefont {Wineland},\
  and\ \citenamefont {Rosenband}}]{Chou2010}%
  \BibitemOpen
\bibfield  {journal} {  }\bibfield  {author} {\bibinfo {author} {\bibfnamefont
  {C.~W.}\ \bibnamefont {Chou}}, \bibinfo {author} {\bibfnamefont {D.~B.}\
  \bibnamefont {Hume}}, \bibinfo {author} {\bibfnamefont {J.~C.~J.}\
  \bibnamefont {Koelemeij}}, \bibinfo {author} {\bibfnamefont {D.~J.}\
  \bibnamefont {Wineland}}, \ and\ \bibinfo {author} {\bibfnamefont
  {T.}~\bibnamefont {Rosenband}},\ }\href {\doibase
  10.1103/PhysRevLett.104.070802} {\bibfield  {journal} {\bibinfo  {journal}
  {Physical Review Letters}\ }\textbf {\bibinfo {volume} {104}},\ \bibinfo
  {pages} {1} (\bibinfo {year} {2010})}\BibitemShut {NoStop}%
\bibitem [{\citenamefont {Hinkley}\ \emph {et~al.}(2013)\citenamefont
  {Hinkley}, \citenamefont {Sherman}, \citenamefont {Phillips}, \citenamefont
  {Schioppo}, \citenamefont {Lemke}, \citenamefont {Beloy}, \citenamefont
  {Pizzocaro}, \citenamefont {Oates},\ and\ \citenamefont
  {Ludlow}}]{Hinkley2013}%
  \BibitemOpen
  \bibfield  {author} {\bibinfo {author} {\bibfnamefont {N.}~\bibnamefont
  {Hinkley}}, \bibinfo {author} {\bibfnamefont {J.~A.}\ \bibnamefont
  {Sherman}}, \bibinfo {author} {\bibfnamefont {N.~B.}\ \bibnamefont
  {Phillips}}, \bibinfo {author} {\bibfnamefont {M.}~\bibnamefont {Schioppo}},
  \bibinfo {author} {\bibfnamefont {N.~D.}\ \bibnamefont {Lemke}}, \bibinfo
  {author} {\bibfnamefont {K.}~\bibnamefont {Beloy}}, \bibinfo {author}
  {\bibfnamefont {M.}~\bibnamefont {Pizzocaro}}, \bibinfo {author}
  {\bibfnamefont {C.~W.}\ \bibnamefont {Oates}}, \ and\ \bibinfo {author}
  {\bibfnamefont {A.~D.}\ \bibnamefont {Ludlow}},\ }\href {\doibase
  10.1126/science.1240420} {\bibfield  {journal} {\bibinfo  {journal}
  {Science}\ }\textbf {\bibinfo {volume} {341}},\ \bibinfo {pages} {1215}
  (\bibinfo {year} {2013})}\BibitemShut {NoStop}%
\bibitem [{\citenamefont {Bloom}\ \emph {et~al.}(2014)\citenamefont {Bloom},
  \citenamefont {Nicholson}, \citenamefont {Williams}, \citenamefont
  {Campbell}, \citenamefont {Bishof}, \citenamefont {Zhang}, \citenamefont
  {Zhang}, \citenamefont {Bromley},\ and\ \citenamefont {Ye}}]{Bloom2014}%
  \BibitemOpen
  \bibfield  {author} {\bibinfo {author} {\bibfnamefont {B.~J.}\ \bibnamefont
  {Bloom}}, \bibinfo {author} {\bibfnamefont {T.~L.}\ \bibnamefont
  {Nicholson}}, \bibinfo {author} {\bibfnamefont {J.~R.}\ \bibnamefont
  {Williams}}, \bibinfo {author} {\bibfnamefont {S.~L.}\ \bibnamefont
  {Campbell}}, \bibinfo {author} {\bibfnamefont {M.}~\bibnamefont {Bishof}},
  \bibinfo {author} {\bibfnamefont {X.}~\bibnamefont {Zhang}}, \bibinfo
  {author} {\bibfnamefont {W.}~\bibnamefont {Zhang}}, \bibinfo {author}
  {\bibfnamefont {S.~L.}\ \bibnamefont {Bromley}}, \ and\ \bibinfo {author}
  {\bibfnamefont {J.}~\bibnamefont {Ye}},\ }\href {\doibase
  10.1038/nature12941} {\bibfield  {journal} {\bibinfo  {journal} {Nature}\
  }\textbf {\bibinfo {volume} {506}},\ \bibinfo {pages} {71} (\bibinfo {year}
  {2014})}\BibitemShut {NoStop}%
\bibitem [{\citenamefont {Huntemann}\ \emph {et~al.}(2016)\citenamefont
  {Huntemann}, \citenamefont {Sanner}, \citenamefont {Lipphardt}, \citenamefont
  {Tamm},\ and\ \citenamefont {Peik}}]{Huntemann2016}%
  \BibitemOpen
  \bibfield  {author} {\bibinfo {author} {\bibfnamefont {N.}~\bibnamefont
  {Huntemann}}, \bibinfo {author} {\bibfnamefont {C.}~\bibnamefont {Sanner}},
  \bibinfo {author} {\bibfnamefont {B.}~\bibnamefont {Lipphardt}}, \bibinfo
  {author} {\bibfnamefont {C.}~\bibnamefont {Tamm}}, \ and\ \bibinfo {author}
  {\bibfnamefont {E.}~\bibnamefont {Peik}},\ }\href {\doibase
  10.1103/PhysRevLett.116.063001} {\bibfield  {journal} {\bibinfo  {journal}
  {Physical Review Letters}\ }\textbf {\bibinfo {volume} {116}},\ \bibinfo
  {pages} {063001} (\bibinfo {year} {2016})}\BibitemShut {NoStop}%
\bibitem [{\citenamefont {Berengut}\ \emph {et~al.}(2010)\citenamefont
  {Berengut}, \citenamefont {Dzuba},\ and\ \citenamefont
  {Flambaum}}]{Berengut2010}%
  \BibitemOpen
  \bibfield  {author} {\bibinfo {author} {\bibfnamefont {J.~C.}\ \bibnamefont
  {Berengut}}, \bibinfo {author} {\bibfnamefont {V.~A.}\ \bibnamefont {Dzuba}},
  \ and\ \bibinfo {author} {\bibfnamefont {V.~V.}\ \bibnamefont {Flambaum}},\
  }\href {\doibase 10.1103/PhysRevLett.105.120801} {\bibfield  {journal}
  {\bibinfo  {journal} {Physical Review Letters}\ }\textbf {\bibinfo {volume}
  {105}},\ \bibinfo {pages} {120801} (\bibinfo {year} {2010})}\BibitemShut
  {NoStop}%
\bibitem [{\citenamefont {Derevianko}\ \emph {et~al.}(2012)\citenamefont
  {Derevianko}, \citenamefont {Dzuba},\ and\ \citenamefont
  {Flambaum}}]{Dzuba2012HCI}%
  \BibitemOpen
  \bibfield  {author} {\bibinfo {author} {\bibfnamefont {A.}~\bibnamefont
  {Derevianko}}, \bibinfo {author} {\bibfnamefont {V.~A.}\ \bibnamefont
  {Dzuba}}, \ and\ \bibinfo {author} {\bibfnamefont {V.~V.}\ \bibnamefont
  {Flambaum}},\ }\href {\doibase 10.1103/PhysRevLett.109.180801} {\bibfield
  {journal} {\bibinfo  {journal} {Physical Review Letters}\ }\textbf {\bibinfo
  {volume} {109}},\ \bibinfo {pages} {180801} (\bibinfo {year}
  {2012})}\BibitemShut {NoStop}%
\bibitem [{\citenamefont {Windberger}\ \emph {et~al.}(2015)\citenamefont
  {Windberger}, \citenamefont {Crespo L\'opez-Urrutia}, \citenamefont {Bekker},
  \citenamefont {Oreshkina}, \citenamefont {Berengut}, \citenamefont {Bock},
  \citenamefont {Borschevsky}, \citenamefont {Dzuba}, \citenamefont {Eliav},
  \citenamefont {Harman}, \citenamefont {Kaldor}, \citenamefont {Kaul},
  \citenamefont {Safronova}, \citenamefont {Flambaum}, \citenamefont {Keitel},
  \citenamefont {Schmidt}, \citenamefont {Ullrich},\ and\ \citenamefont
  {Versolato}}]{Windberger2015}%
  \BibitemOpen
  \bibfield  {author} {\bibinfo {author} {\bibfnamefont {A.}~\bibnamefont
  {Windberger}}, \bibinfo {author} {\bibfnamefont {J.~R.}\ \bibnamefont {Crespo
  L\'opez-Urrutia}}, \bibinfo {author} {\bibfnamefont {H.}~\bibnamefont
  {Bekker}}, \bibinfo {author} {\bibfnamefont {N.~S.}\ \bibnamefont
  {Oreshkina}}, \bibinfo {author} {\bibfnamefont {J.~C.}\ \bibnamefont
  {Berengut}}, \bibinfo {author} {\bibfnamefont {V.}~\bibnamefont {Bock}},
  \bibinfo {author} {\bibfnamefont {A.}~\bibnamefont {Borschevsky}}, \bibinfo
  {author} {\bibfnamefont {V.~A.}\ \bibnamefont {Dzuba}}, \bibinfo {author}
  {\bibfnamefont {E.}~\bibnamefont {Eliav}}, \bibinfo {author} {\bibfnamefont
  {Z.}~\bibnamefont {Harman}}, \bibinfo {author} {\bibfnamefont
  {U.}~\bibnamefont {Kaldor}}, \bibinfo {author} {\bibfnamefont
  {S.}~\bibnamefont {Kaul}}, \bibinfo {author} {\bibfnamefont {U.~I.}\
  \bibnamefont {Safronova}}, \bibinfo {author} {\bibfnamefont {V.~V.}\
  \bibnamefont {Flambaum}}, \bibinfo {author} {\bibfnamefont {C.~H.}\
  \bibnamefont {Keitel}}, \bibinfo {author} {\bibfnamefont {P.~O.}\
  \bibnamefont {Schmidt}}, \bibinfo {author} {\bibfnamefont {J.}~\bibnamefont
  {Ullrich}}, \ and\ \bibinfo {author} {\bibfnamefont {O.~O.}\ \bibnamefont
  {Versolato}},\ }\href {\doibase 10.1103/PhysRevLett.114.150801} {\bibfield
  {journal} {\bibinfo  {journal} {Physical Review Letters}\ }\textbf {\bibinfo
  {volume} {114}},\ \bibinfo {pages} {150801} (\bibinfo {year}
  {2015})}\BibitemShut {NoStop}%
\bibitem [{\citenamefont {Flambaum}\ and\ \citenamefont
  {Kozlov}(2007)}]{Flambaum2007Mol}%
  \BibitemOpen
  \bibfield  {author} {\bibinfo {author} {\bibfnamefont {V.~V.}\ \bibnamefont
  {Flambaum}}\ and\ \bibinfo {author} {\bibfnamefont {M.~G.}\ \bibnamefont
  {Kozlov}},\ }\href {\doibase 10.1103/PhysRevLett.99.150801} {\bibfield
  {journal} {\bibinfo  {journal} {Physical Review Letters}\ }\textbf {\bibinfo
  {volume} {99}},\ \bibinfo {pages} {150801} (\bibinfo {year}
  {2007})}\BibitemShut {NoStop}%
\bibitem [{\citenamefont {Zelevinsky}\ \emph {et~al.}(2008)\citenamefont
  {Zelevinsky}, \citenamefont {Kotochigova},\ and\ \citenamefont
  {Ye}}]{Ye2008Mol}%
  \BibitemOpen
  \bibfield  {author} {\bibinfo {author} {\bibfnamefont {T.}~\bibnamefont
  {Zelevinsky}}, \bibinfo {author} {\bibfnamefont {S.}~\bibnamefont
  {Kotochigova}}, \ and\ \bibinfo {author} {\bibfnamefont {J.}~\bibnamefont
  {Ye}},\ }\href {\doibase 10.1103/PhysRevLett.100.043201} {\bibfield
  {journal} {\bibinfo  {journal} {Physical Review Letters}\ }\textbf {\bibinfo
  {volume} {100}},\ \bibinfo {pages} {043201} (\bibinfo {year}
  {2008})}\BibitemShut {NoStop}%
\bibitem [{\citenamefont {DeMille}\ \emph {et~al.}(2008)\citenamefont
  {DeMille}, \citenamefont {Sainis}, \citenamefont {Sage}, \citenamefont
  {Bergeman}, \citenamefont {Kotochigova},\ and\ \citenamefont
  {Tiesinga}}]{DeMille2007Mol}%
  \BibitemOpen
  \bibfield  {author} {\bibinfo {author} {\bibfnamefont {D.}~\bibnamefont
  {DeMille}}, \bibinfo {author} {\bibfnamefont {S.}~\bibnamefont {Sainis}},
  \bibinfo {author} {\bibfnamefont {J.}~\bibnamefont {Sage}}, \bibinfo {author}
  {\bibfnamefont {T.}~\bibnamefont {Bergeman}}, \bibinfo {author}
  {\bibfnamefont {S.}~\bibnamefont {Kotochigova}}, \ and\ \bibinfo {author}
  {\bibfnamefont {E.}~\bibnamefont {Tiesinga}},\ }\href {\doibase
  10.1103/PhysRevLett.100.043202} {\bibfield  {journal} {\bibinfo  {journal}
  {Physical Review Letters}\ }\textbf {\bibinfo {volume} {100}},\ \bibinfo
  {pages} {043202} (\bibinfo {year} {2008})}\BibitemShut {NoStop}%
\bibitem [{\citenamefont {Peik}\ and\ \citenamefont {Tamm}(2003)}]{Peik2003}%
  \BibitemOpen
  \bibfield  {author} {\bibinfo {author} {\bibfnamefont {E.}~\bibnamefont
  {Peik}}\ and\ \bibinfo {author} {\bibfnamefont {C.}~\bibnamefont {Tamm}},\
  }\href {\doibase 10.1209/epl/i2003-00210-x} {\bibfield  {journal} {\bibinfo
  {journal} {Europhysics Letters}\ }\textbf {\bibinfo {volume} {61}},\ \bibinfo
  {pages} {181} (\bibinfo {year} {2003})}\BibitemShut {NoStop}%
\bibitem [{\citenamefont {Flambaum}(2006)}]{Flambaum2006Th}%
  \BibitemOpen
  \bibfield  {author} {\bibinfo {author} {\bibfnamefont {V.~V.}\ \bibnamefont
  {Flambaum}},\ }\href {\doibase 10.1103/PhysRevLett.97.092502} {\bibfield
  {journal} {\bibinfo  {journal} {Physical Review Letters}\ }\textbf {\bibinfo
  {volume} {97}},\ \bibinfo {pages} {092502} (\bibinfo {year}
  {2006})}\BibitemShut {NoStop}%
\bibitem [{\citenamefont {Kazakov}\ \emph {et~al.}(2012)\citenamefont
  {Kazakov}, \citenamefont {Litvinov}, \citenamefont {Romanenko}, \citenamefont
  {Yatsenko}, \citenamefont {Romanenko}, \citenamefont {Schreitl},
  \citenamefont {Winkler},\ and\ \citenamefont {Schumm}}]{Kazakov2012}%
  \BibitemOpen
  \bibfield  {author} {\bibinfo {author} {\bibfnamefont {G.~A.}\ \bibnamefont
  {Kazakov}}, \bibinfo {author} {\bibfnamefont {A.~N.}\ \bibnamefont
  {Litvinov}}, \bibinfo {author} {\bibfnamefont {V.~I.}\ \bibnamefont
  {Romanenko}}, \bibinfo {author} {\bibfnamefont {L.~P.}\ \bibnamefont
  {Yatsenko}}, \bibinfo {author} {\bibfnamefont {A.~V.}\ \bibnamefont
  {Romanenko}}, \bibinfo {author} {\bibfnamefont {M.}~\bibnamefont {Schreitl}},
  \bibinfo {author} {\bibfnamefont {G.}~\bibnamefont {Winkler}}, \ and\
  \bibinfo {author} {\bibfnamefont {T.}~\bibnamefont {Schumm}},\ }\href
  {\doibase 10.1088/1367-2630/14/8/083019} {\bibfield  {journal} {\bibinfo
  {journal} {New Journal of Physics}\ }\textbf {\bibinfo {volume} {14}},\
  \bibinfo {pages} {083019} (\bibinfo {year} {2012})}\BibitemShut {NoStop}%
\bibitem [{\citenamefont {Stadnik}\ and\ \citenamefont
  {Flambaum}(2015{\natexlab{b}})}]{Stadnik2015Laser}%
  \BibitemOpen
  \bibfield  {author} {\bibinfo {author} {\bibfnamefont {Y.~V.}\ \bibnamefont
  {Stadnik}}\ and\ \bibinfo {author} {\bibfnamefont {V.~V.}\ \bibnamefont
  {Flambaum}},\ }\href {\doibase 10.1103/PhysRevLett.114.161301} {\bibfield
  {journal} {\bibinfo  {journal} {Physical Review Letters}\ }\textbf {\bibinfo
  {volume} {114}},\ \bibinfo {pages} {161301} (\bibinfo {year}
  {2015}{\natexlab{b}})}\BibitemShut {NoStop}%
\end{thebibliography}%

\end{document}